\documentclass[journal]{IEEEtran}
\ifCLASSINFOpdf
\else
\fi

\usepackage[utf8]{inputenc}

\usepackage{graphicx}

\usepackage[hidelinks]{hyperref}

\usepackage[nomain, toc, acronym]{glossaries}

\usepackage{varioref}

\usepackage{csquotes}
\usepackage[backend=bibtex,style=numeric,sorting=none,natbib=true]{biblatex}

\usepackage[binary-units=true]{siunitx}

\usepackage{multirow}
\usepackage{booktabs}
\usepackage{threeparttable}

\usepackage{commath}

\usepackage{amsmath}

\usepackage{interval}

\usepackage{microtype}
\usepackage{xspace}
\def\mycommand{\bgroup\obeyspaces\mycommandaux}
\def\mycommandaux#1{\mycommandauxii #1\relax\relax\egroup}
\def\mycommandauxii#1{%
	\ifx\relax#1\else \ifcat#1\@sptoken{} \expandafter\expandafter\expandafter\mycommandauxii\else
	\ifnum`#1=\uccode`#1 {\normalsize #1}\else {\footnotesize \uppercase{#1}}\fi \expandafter\expandafter\expandafter\mycommandauxii\expandafter\fi\fi}
\newcommand{\sync}{\mbox{\textls[50]{\mycommand{sync}}}\xspace}
\newcommand{\followup}{\mbox{\textls[50]{\mycommand{follow{ }up}}}\xspace}
\newcommand{\delayreq}{\mbox{\textls[50]{\mycommand{delay{ }request}}}\xspace}
\newcommand{\delayresp}{\mbox{\textls[50]{\mycommand{delay{ }response}}}\xspace}
\newcommand{\synce}{\mbox{\textls[50]{\mycommand{sync{ }early}}}\xspace}
\newcommand{\syncl}{\mbox{\textls[50]{\mycommand{sync{ }late}}}\xspace}
\newcommand{\delayreql}{\mbox{\textls[50]{\mycommand{delay{ }request{ }late}}}\xspace}
\newcommand{\delayreqe}{\mbox{\textls[50]{\mycommand{delay{ }request{ }early}}}\xspace}
\newcommand{\announce}{\mbox{\textls[50]{\mycommand{announce}}}\xspace}
\def\mycommandsmall{\bgroup\obeyspaces\mycommandsmallaux}
\def\mycommandsmallaux#1{\mycommandsmallauxii #1\relax\relax\egroup}
\def\mycommandsmallauxii#1{%
	\ifx\relax#1\else \ifcat#1\@sptoken{} \expandafter\expandafter\expandafter\mycommandsmallauxii\else
	\ifnum`#1=\uccode`#1 {\scriptsize #1}\else {\tiny \uppercase{#1}}\fi \expandafter\expandafter\expandafter\mycommandsmallauxii\expandafter\fi\fi}
\newcommand{\syncsmall}{\mbox{\textls[50]{\mycommandsmall{sync}}}\xspace}

\newcommand{\delayreqsmall}{\mbox{\textls[50]{\mycommandsmall{delay{ }request}}}\xspace}
\newcommand{\delayrespsmall}{\mbox{\textls[50]{\mycommandsmall{delay{ }response}}}\xspace}

\graphicspath{{./figures/}}

\AtBeginBibliography{\small}

\clearpage{}%
\newacronym{cdn}{CDN}{Content Delivery Network}
\newacronym{osi}{OSI}{Open Systems Interconnection}
\newacronym{ip}{IP}{Internet Protocol}
\newacronym{alm}{ALM}{Application Layer Multicast}
\newacronym{wampacs}{WAMPACS}{Wide Area Monitoring, Protection, and Control System}
\newacronym{mac}{MAC}{Message Authentication Code}
\newacronym{ptp}{PTP}{Precision Time Protocol}
\newacronym{ntp}{NTP}{Network Time Protocol}
\newacronym{ietf}{IETF}{Internet Engineering Task Force}
\newacronym{ami}{AMI}{Advanced Metering Infrastructure}
\newacronym{wams}{WAMS}{Wide Area Monitoring System}
\newacronym{pmu}{PMU}{Phasor Measurement Unit}
\newacronym{p2p}{P2P}{Peer to Peer}
\newacronym{sips}{SIPS}{System Integrity Protection System}
\newacronym{sps}{SPS}{System Protection System}
\newacronym{mmog}{MMOG}{Massively Multiplayer Online Game}
\newacronym{mdns}{mDNS}{Multicast DNS}
\newacronym{iot}{IoT}{Internet of Things}
\newacronym{pki}{PKI}{Public Key Infrastructure}
\newacronym{mts}{MTS}{Multiple-Time Signature}
\newacronym{ecc}{ECC}{Elliptic Curve Cryptography}
\newacronym{ots}{OTS}{One-Time Signature}
\newacronym{dos}{DoS}{Denial of Service}
\newacronym{atm}{ATM}{Asynchronous Transfer Mode}
\newacronym{tls}{TLS}{Transport Layer Security}
\newacronym{tesla}{TESLA}{Timed Efficient Stream Loss-tolerant Authentication}
\newacronym{nts}{NTS}{Network Time Security}
\newacronym{tvhors}{TV-HORS}{Time Valid Hash to Obtain Random Subsets}
\newacronym{sntp}{SNTP}{Simple Network Time Protocol}
\newacronym{ida}{IDA}{Information Dispersal Algorithm}
\newacronym{dsm}{DSM}{Demand Side Management}
\newacronym{dr}{DR}{Demand Response}
\newacronym{pps}{PPS}{Pulses per Second}
\newacronym{gps}{GPS}{Global Positioning System}
\newacronym{cots}{COTS}{Commercial Off-The-Shelf}
\newacronym{rlh}{RLH}{Receiver driven Layered Hash-chaining}
\newacronym{tlv}{TLV}{Type Length Value}
\newacronym{ids}{IDS}{Intrusion Detection System}
\newacronym{mitm}{MITM}{Man in the Middle}
\newacronym{mq}{MQ}{Multivariate-Quadratic}
\newacronym{nic}{NIC}{Network Interface Card}
\newacronym{wan}{WAN}{Wide Area Network}
\newacronym{lan}{LAN}{Local Area Network}
\newacronym{utc}{UTC}{Universal Time Coordinated}
\newacronym{rtd}{RTD}{Round-Trip Delay}
\newacronym{m2m}{M2M}{Machine-to-Machine}
\newacronym{esp}{ESP}{Encapsulating Security Payload}
\newacronym{tfc}{TFC}{Traffic Flow Confidentiality}
\newacronym{owd}{OWD}{One-Way Delay}\clearpage{}%

\begin{document}
	\title{Encryption is Futile: Delay Attacks on High-Precision Clock Synchronization}

	\author{\IEEEauthorblockN{Robert~Annessi, Joachim~Fabini, Felix~Iglesias, and~Tanja~Zseby}\\
		\IEEEauthorblockA{Institute of Telecommunications\\
			TU Wien, Austria\\
			Email: firstname.lastname@nt.tuwien.ac.at}}

	\markboth
	{R. Annessi, J. Fabini, F.Iglesias, and T. Zseby: Encryption is Futile: Delay Attacks on High-Precision Clock Synchronization}
	{R. Annessi, J. Fabini, F.Iglesias, and T. Zseby: Encryption is Futile: Delay Attacks on High-Precision Clock Synchronization}

	\maketitle

	\begin{abstract}
		Clock synchronization has become essential to modern societies since many critical infrastructures depend on a precise notion of time. 
		This paper analyzes security aspects of high-precision clock synchronization protocols, particularly their alleged protection against delay attacks when clock synchronization traffic is encrypted using  
		standard network security protocols such as IPsec, MACsec, or TLS. 
		We use the \gls{ptp}, the most widely used protocol for high-precision clock synchronization, to demonstrate that statistical traffic analysis can identify properties that support selective message delay attacks even for encrypted traffic. 
		We furthermore identify a fundamental conflict in secure clock synchronization between the need of deterministic traffic to improve precision and the need to obfuscate traffic in order to mitigate delay attacks. 
		
		A theoretical analysis of clock synchronization protocols isolates the characteristics that make these protocols vulnerable to delay attacks and argues that such attacks cannot be prevented entirely but only be mitigated. 
		Knowledge of the underlying communication network in terms of one-way delays and knowledge on physical constraints of these networks can help to compute guaranteed maximum bounds for slave clock offsets. These bounds are essential for detecting delay attacks and minimizing their impact. 
		In the general case, however, the precision that can be guaranteed in adversarial settings is orders of magnitude lower than required for high-precision clock synchronization in critical infrastructures, which, therefore, must not rely on a precise notion of time when using untrusted networks. 

	\end{abstract}
	\glsreset{ptp}

	\IEEEpeerreviewmaketitle

	\section{Introduction}
	\label{sec:intro}

\IEEEPARstart{C}{lock} synchronization protocols have become an essential building block of numerous applications that rely on a precise notion of time. 
The deployment of clock synchronization for controlling system clocks of critical applications in telecommunication, industrial automation, financial markets, avionics, or energy distribution has increased the dependency of critical infrastructures on clocks synchronized with increasingly high precision. 
Many processes, especially in measurement, control, telecommunications, industrial and financial applications are not only sensitive to errors in the value domain but also to errors in the time domain. 
Examples for strict dependencies on precise time are safety-critical applications in the Smart Grid, which require an accuracy of \numrange{1}{100}~\si{\micro\second} (\num{10}~\si{\micro\second} in case of current differential line protection with high fault current sensitivity~\cites{61850-90-1}{C37.118.1-2011}) or MiFID II in the financial sector, requiring an accuracy of up to \SI{100}{\micro\second}~\cites{mfid}{mfida}{mfidb}.
Cellular networks also have strong requirements for synchronized clocks ($\le$ \SI{1}{\micro\second})~\cite{cellularclock}. 
Errors in clock synchronization can lead to wrong timings and may therefore originate faulty sensor reports, endanger control decisions, and adversely affect the overall functionality of a wide range of (critical) services that depend on accurate time. 

In recent years, security of clock synchronization received increased attention as various attacks on clock synchronization protocols (and countermeasures) were proposed. 
For this reason, clock synchronization protocols need to be secured whenever used outside of fully trusted network environments. 
Clock synchronization protocols are specifically susceptible to delay attacks since the times when messages are sent and received have an actual effect on the receiver's notion of time. 
Delay attacks specifically can degrade the precision of clock synchronization such that applications depending on it may malfunction. 
Such malfunctioning is endangering critical infrastructures that increasingly depend on a precise notion of time. 

Encrypting and signing clock synchronization messages (with IPsec~\cite{rfc6071}, MACsec~\cite{macsec}, or TLS~\cites{rfc8446}) is commonly recommended in order to provide security against a wealth of network-based attacks. 
In this paper, we show that encryption alone cannot provide sufficient security against delay attacks on high-precision clock synchronization, which highlights the insecurity of numerous applications (including critical infrastructures) that depend on a precise notion of time. 
We present a fundamental limitation of clock synchronization over untrusted networks that is independent of the actual clock synchronization protocol and communication network and even holds when the communication is (supposedly) secured.

\subsection{Contributions}
The main contributions of this paper are the following:
\begin{itemize}
\item \textbf{\gls{ptp} traffic detection by statistical analysis}: We present a method for statistical \textit{traffic analysis} of \gls{ptp}'s 2-step mode that allows to identify \gls{ptp} traffic even if it is encrypted and transmitted together with other traffic. We show that some statistical properties (length, timing, and direction) are sufficient to identify \gls{ptp} traffic and the particular \gls{ptp} message types with very high probability --- even when the traffic is encrypted. 

\item \textbf{Selective message delay attacks on encrypted traffic}: Using the properties identified in the traffic analysis, we design and implement \textit{selective message delay attacks} on encrypted traffic. 
This shows that delay attacks can be conducted on \gls{ptp} even when it is (supposedly) secured with state-of-the-art network encryption schemes such as IPsec, MACsec, or TLS. 
\item \textbf{Countermeasures}: We discuss countermeasures against traffic analysis (length padding, timing randomization), selective message delay attacks (strict replay protection), and asymmetric link delay attacks (\gls{owd} limits). 
\item \textbf{General clock synchronization limitations in adversarial settings}: We analyze the theoretical foundations that make clock synchronization protocols susceptible to delay attacks and identify the main vulnerability to be the delay compensation mechanism, which is necessary to achieve high precision. 
This leads us to \textit{asymmetric link delay attacks} that cannot be prevented by encryption. 
While additional knowledge of the underlying communication network can bound the maximum impact of delay attacks, we argue that clock synchronization protocols cannot be designed in a way that prevents asymmetric link delay attacks entirely. 
\end{itemize}
The resulting bounds, which we found based on our analysis, do not satisfy requirements with respect to high-precision clock synchronization. Therefore, high precision can only be guaranteed on trusted networks.  	
	\section{Background}
	\label{sec:back}

Each clock has a natural drift caused by the non-ideality of physical oscillators such an oscillator's frequency affected by temperature. 
The aim of clock synchronization is to keep clocks within acceptable boundaries. 
Currently, there are two widely used technologies for high-precision clock synchronization: (1) the satellite-based \gls{gps} for \gls{pps} synchronization to global \gls{utc}, and \gls{ptp} targeted for network-based clock synchronization to a local clock. %
The main disadvantages of \gls{gps} \gls{pps} are that it is operated by a single entity (the US Air Force) and that it requires free view on at least four satellites\footnote{View to four \gls{gps} satellites is the requirement for \gls{gps}-\gls{pps} signal generation in the general case. 
However, from a theoretical point of view, one single satellite in view is sufficient whenever the exact \gls{gps} antenna position is known}, which might be difficult to get (in data centers for example). 
Moreover, the (public) \gls{gps} signal is not secured so that it may be spoofed with reasonable effort~\cite{gpsspoof}. 
Due to the missing backchannel to satellites, \gls{gps} is conceptionally different to the \gls{ntp} and \gls{ptp}. 
In this paper, we focus on \gls{ptp} because it can achieve extremely high precision, is widely used, and is more similar to other clock synchronization protocols such as \gls{ntp}. 
Most of the findings in this paper, however, are applicable to other clock synchronization protocols as well. 

\gls{ptp}~\cite{ptp} (or IEEE 1588) is designed to provide high-precision clock synchronization in the order of microseconds or even sub-microseconds. 
\gls{ptp} achieves such high precision mainly by reducing the impact of delay variation in to the hosts' operating systems. 
For this purpose, \glspl{nic} with dedicated \gls{ptp} hardware support can precisely timestamp \gls{ptp} messages at send and receive times in order to eliminate software uncertainty. 
\gls{ptp} assumes trustworthy and somewhat deterministic networks with low latency that are entirely under control of the operator. 
It can be run over a \gls{lan} via Ethernet or over a \gls{wan} via UDP. 

New applications that require increasingly precise clock synchronization lead to \gls{ptp} being deployed also in critical infrastructures and other areas that \gls{ptp} has not been designed for specifically. 
Base stations in mobile networks that depend on highly accurate time are a prominent example. 
Originally, manufacturers used \gls{gps} \gls{pps} to synchronize clocks. 
Nowadays manufacturers also offer \gls{ptp} as an alternative that is widely used. 
Given the fact that base stations typically use \glspl{wan} to connect to their core network, employing \gls{ptp} contradicts the original assumptions of being used within \glspl{lan}. 
In this way, new attack vectors are created such as the delay attacks presented in this paper. 
More examples include but are not limited to areas such as Smart Grids or financial applications.

\subsection{The Two Phases in Clock Synchronization}
Clock synchronization algorithms aim at synchronizing a slave clock to a master clock by exchanging timestamped messages over packet-switched networks. 
In this paper we assume that the master's clock is accurate and reliable. 
Details on how the master implements and accesses such a reliable clock are out of scope of this paper.  
Network-based clock synchronization protocols depend on two distinct phases that will be discussed below: (a) clock offset measurement and (b) delay measurement. 

\begin{figure}
	\centering
	\includegraphics[width=0.9\linewidth]{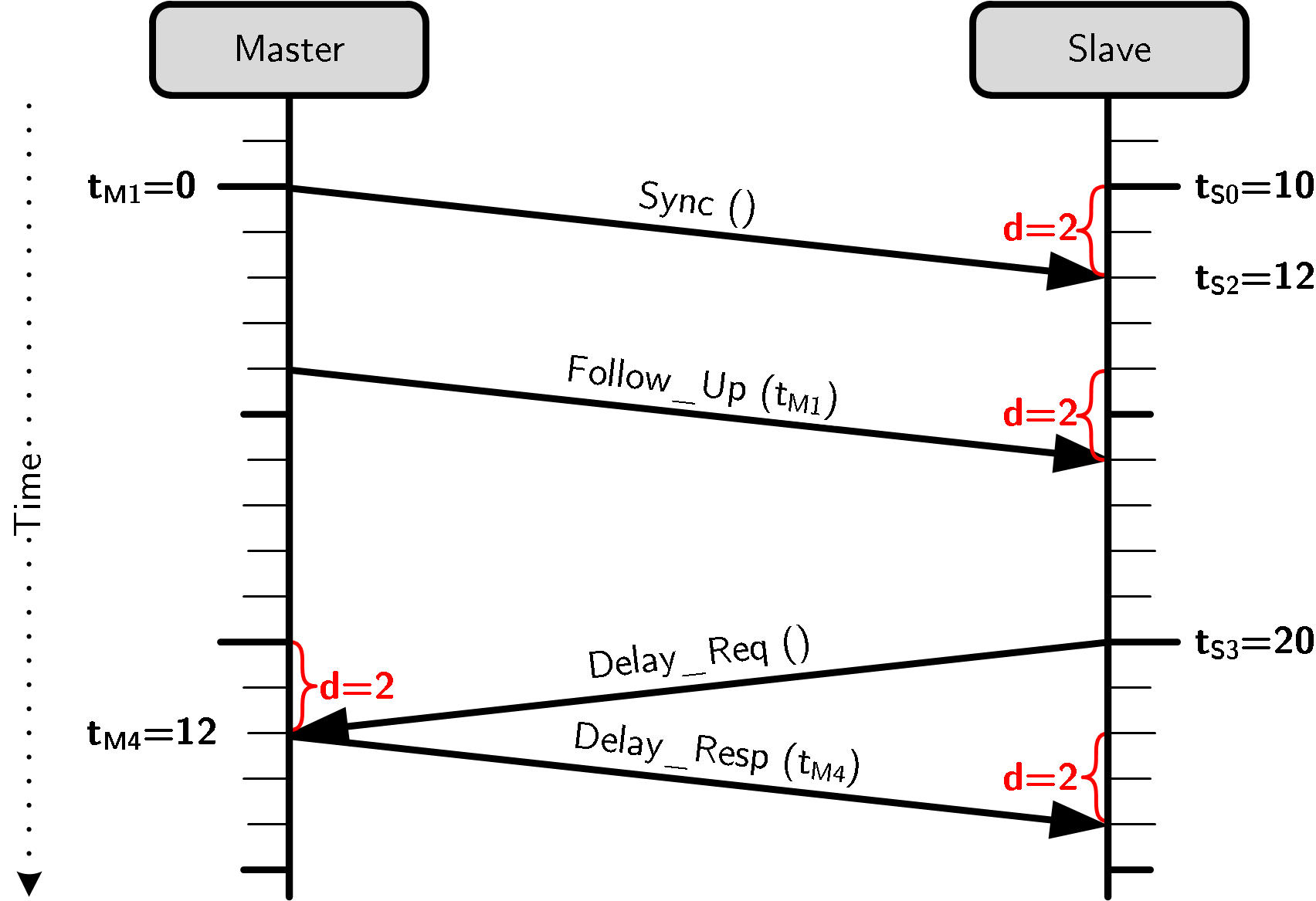}
	\caption{2-step clock synchronization with deterministic and symmetric delays and initial clock offset of \num{10} time units.}
	\label{fig:ptp_}
\end{figure}

\paragraph{Clock Offset Measurement Phase}
The goal of the clock offset measurement phase is to calculate the relative difference between the slave and master clocks. 
Clock offset can either be measured in a single message (\num{1}-step mode) supported by both \gls{ntp} and \gls{ptp} or in two messages (2-step mode) supported by \gls{ptp}. 
In any case, the master sends a \sync message to the slaves, and the slave records the transmitting time $t_{M1}$. 
In \num{1}-step mode the \sync message contains the transmitting timestamp ($t_{M1}$), in two-step mode the \sync message is just used as a marker, and the \followup message contains the exact point in time when the \sync message left the master ($t_{M1}$). 
This way, higher precision may be achieved. 
Fig.~\ref{fig:ptp_} depicts the \num{2}-step clock offset measurement and delay measurement. 

\paragraph{Delay Measurement Phase}
The \sync and \followup messages are subject to various delay. 
These delays are added to (and therefore negatively affect) the measured clock offset. 
The overall delay consists of transmission delays, queuing delays, processing delays, and propagation delays, which themselves consist of constant and stochastic parts\footnote{There is some constant delay for a given route and message size and stochastic delay that mainly depends on other traffic and states of network devices.}. 
The goal of the delay measurement phase is to measure the overall delay and to subtract it from the measured clock offset in order to derive the actual clock offset as precisely as possible. 

In \gls{ptp} a delay measurement consists of two messages, one sent from the slave to the master (\delayreq) where the slave records the transmitting time $t_{S3}$ and a subsequent message from the master to the slave (\delayresp) that includes the time instant when the \delayreq was received at the master ($t_{M4}$). 
Eventually, the slave knows four timestamps: $t_{M1}$, $t_{S2}$, $t_{S3}$, and $t_{M4}$. 
The slave calculates the network \gls{rtd} by measuring the \glspl{owd} in both directions (see Eq.~\ref{eq:rtd}). 
The \gls{rtd} is calculated as the sum of the delay from master to slave ($t_{S2} - t_{M1})$ and the delay from slave to master ($t_{M4} - t_{S3}$). 
The \gls{owd} are approximated as $\frac{RTD}{2}$, assuming symmetric \glspl{owd}. 

\begin{equation}
\mathit{RTD} = t_{S2} - t_{M1} + t_{M4} - t_{S3}
\label{eq:rtd}
\end{equation}

\subsection{Clock Offset Calculation}
For simplicity reasons, in the following example we assume that master and slave clocks do not drift. 
The initial clock offset is \num{10} time units such that the local timestamp $t_{M1} = 0$ on the master corresponds to the local timestamp $t_{S0} = 10$ on the slave (from an external observer's point of view). 
The \glspl{owd} are $2$ time units in each direction. 
The slave calculates its clock offset to the master according to Eq.~\eqref{eq:offset}. 
\begin{equation}
\mathit{offset} = t_{S2} - t_{M1} - \frac{\mathit{RTD}}{2}
\label{eq:offset}
\end{equation}
Eq.~\eqref{eq:offset} consists of the uncorrected clock offset ($t_{S2} - t_{M1}$) corrected with the \gls{owd} that is approximated by halving the \gls{rtd} (Eq.~\ref{eq:rtd}). 
In this specific example, the slave calculates the \gls{owd} as \num{2} and the clock offset as \num{10}. 
Now the slave knows that its clock is \num{10} time units ahead of the master and can adjust accordingly. 
In real-world scenario, all physical clocks are subject to drift so that the process of offset correction needs to be run repeatedly in order to achieve a common notion of time. 
 	
	\section{Threat Model}
	\label{sec:threat}

The goal of the adversary is to make slaves adhere to false clock values or to degrade the precision of the clock synchronization to such an extent that it cannot be considered high-precision anymore. 
We define \textit{high-precision} that for every time instant $i: \vert t_{Si} - t_{Mi} \lvert \le $ \SI{10}{\micro\second} holds, and the goal of the adversary is to disturb this synchronicity of the master and slave clocks at some instant. 

We assume that the adversary is in a privileged \gls{mitm} position in the network, by having gained access to a network node or a link of the communication path, or by conducting an ARP poisoning attack for example. 
The adversary can, therefore, selectively manipulate, capture, and delay any packet on the communication network, in particular any clock synchronization packet that master and slave exchange. 
We assume the communication between master and slaves to be confidential and integrity-protected such that the adversary can neither read nor modify the communication in the value domain but can only modify the time domain. 
The computational power of the adversary is limited but not necessarily bound to that of the master or the slaves. 
I.e., the adversary can use more powerful devices and larger storage. 
 	
	\section{Delay Attacks}
	\label{sec:delayattacks}

Whenever clock synchronization messages are neither encrypted nor integrity-protected, an adversary can attack clock synchronization protocols in the value domain, i.e., the adversary modifies the timestamp values included in the messages. 
This is a well-studied field and various countermeasures have been proposed to secure clock synchronization against attacks in the value domain such as encrypting the communication (with MACsec, IPsec, or TLS) or employing digital signatures to ensure the integrity and authenticity of the communication. 

To conduct an attack in the time domain an adversary intercepts clock synchronization messages and delays them artificially for some time before forwarding\footnotemark. 
The maliciously introduced delay can be constant, variable, or even random, and the slave clock can be manipulated this way~\cite{rfc7384}. 
Since the clock synchronization protocol has no information about the underlying communication network, one fundamental prerequisite and assumption of \gls{ptp} is symmetric network delay between master and slave. 
I.e., the \gls{owd} from master to slave is identical to the \gls{owd} from slave to master. 
Delay attacks exploit this assumption of symmetric delay by maliciously introducing asymmetry such that the slave synchronizes to an inaccurate time. 

\footnotetext{Theoretically delay attacks can also be conducted by accelerating messages (instead of delaying them). 
To conduct such acceleration attack, the adversary needs to delay all messages by default and selectively forward some messages with less delay in order to achieve an acceleration effect.
Another option to accelerate messages would be to route them through faster paths if the attacker has this option.}

The goal of delay attacks is to maliciously manipulate one of the two messages that are crucial to clock offset measurement and delay measurement: \sync and \delayreq. 
While other network-based attacks are important as well, delay attacks and especially countermeasures against delay attacks on clock synchronization have not been studied in required depth yet. 
This paper focuses on this gap in research --- delay attacks and their impact on clock synchronization's precision.

As we will show in Section~\ref{sec:selectiveattack} and~\ref{sec:asymmattack}, delay attacks are feasible despite security measures in place (i.e., traffic being encrypted and integrity-protected). 
Encrypting traffic is not sufficient, mainly because successful verification of a message's integrity only certifies the correctness of the (sending time reported in the) message but not of its effective propagation time through the network~\cite{katz_digital_2010}. 
For this reason, the delay attacks can also be conducted on encrypted and integrity-protected traffic. 
The assumption of symmetric \glspl{owd} is essential to the delay attacks presented in this paper as those exploit non-deterministic delays in communication networks. 

\begin{figure}
	\centering
	\includegraphics[width=0.9\linewidth]{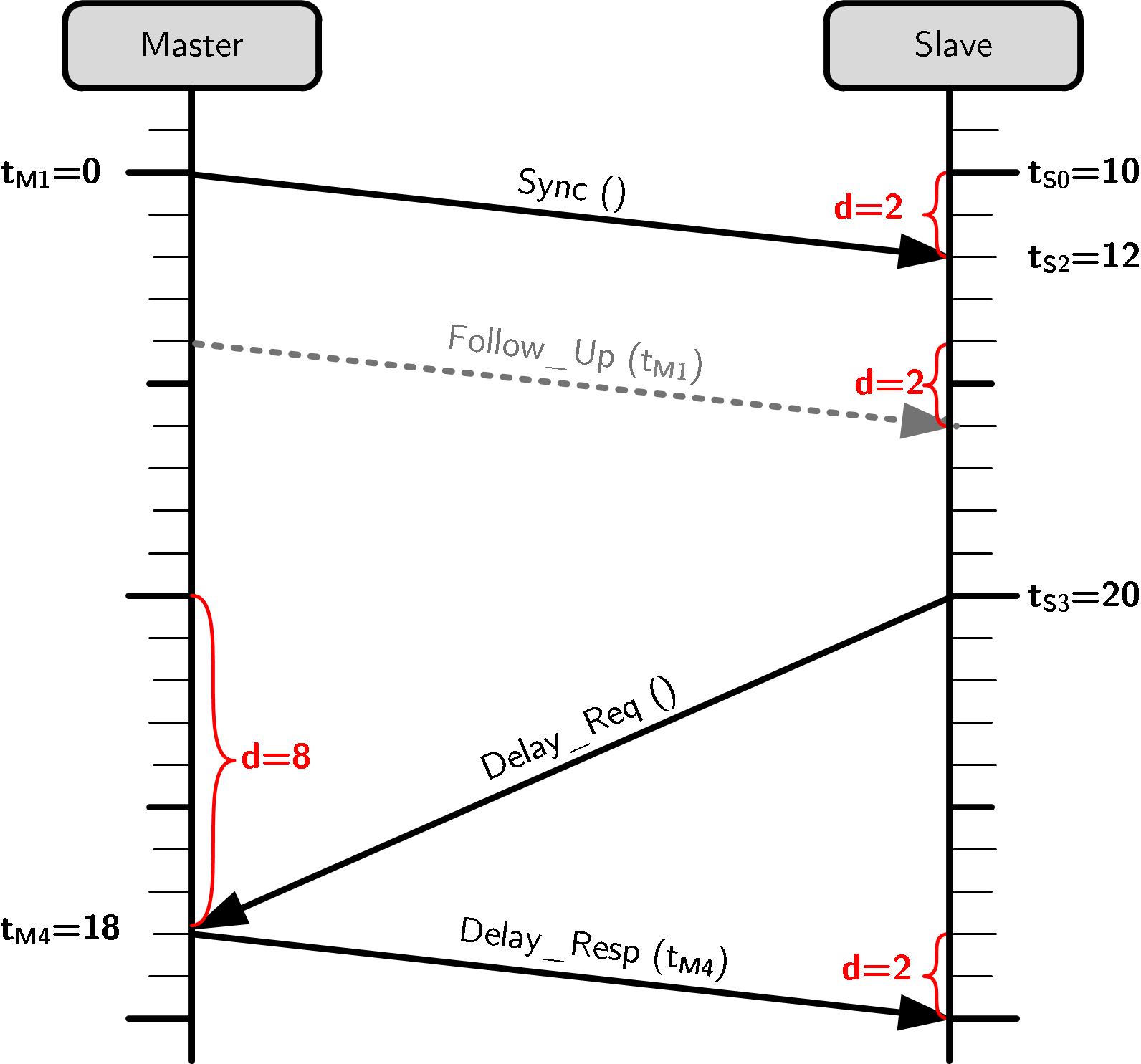}
	\caption{PTP clock synchronization with delayed \delayreqsmall message.}
	\label{fig:ptp_asym_delay}
\end{figure}

\begin{figure}
	\centering
	\includegraphics[width=0.9\linewidth]{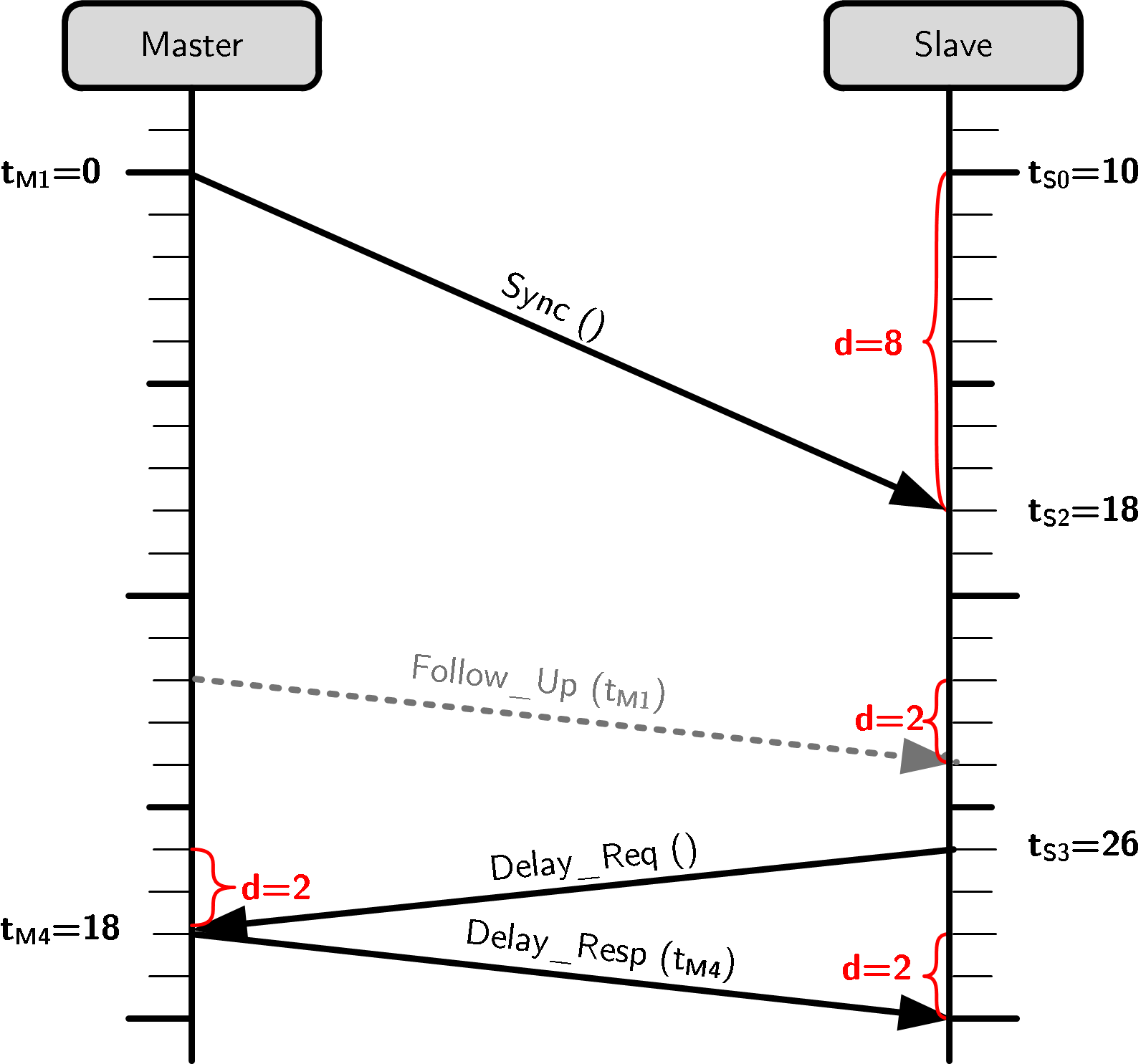}
	\caption{PTP clock synchronization with delayed \syncsmall message.}
	\label{fig:ptp_asym_sync}
\end{figure}

Asymmetric \glspl{owd} have a direct effect on the precision of clock synchronization. 
The effect of asymmetric \glspl{owd} on clock synchronization is depicted in Fig.~\ref{fig:ptp_asym_delay} and~\ref{fig:ptp_asym_sync}. 
In Fig.~\vref{fig:ptp_asym_delay}, the messages from master to slave take \num{2} time units while those from slave to master take \num{8}. 
The slave therefore calculates the clock offset according to Eq.~\ref{eq:offset} as \num{7}, which is off from the actual offset (\num{10}). 
In this example, the slave clock remains \num{3} time units ahead of the master's because of the asymmetric delay in the \delayreq message. 
If the \sync message, on the other hand, takes longer than the \delayreq message (Fig.~\ref{fig:ptp_asym_sync}), the slave miscalculates the clock offset (again according to Eq.~\ref{eq:offset}) as \num{13} so that the slave corrects its clock too much and its clock is then behind the master's. 
Such asymmetric delays are an inherent part of packet-switched communication networks as transmission delays, propagation delays, queuing delays, and processing delays are never entirely symmetric\footnote{Sometimes, asymmetry is even intentional like with Ethernet where cables are asymmetric by design to reduce far end crosstalk~\cite{6645332}.}. 

	\section{Related Work}
	\label{sec:related}

\citeauthor{tsang2005security} where the first to analyze the security of \gls{ptp} in \citeyear{tsang2005security}~\cites{tsang2005security}{tsang_security_2006}. 
They describe delay attacks briefly and propose to average delay measurements as a countermeasure. 
We argue that averaging delay measurements also averages the malicious delay such that the attack is only barely mitigated. 
The attack has even full effect as soon as the duration of the attack is longer than the averaging interval\footnote{And if the averaging interval is too long, the clock synchronization precision may be negatively affected (since reacting on changes in the \gls{rtd} would take too long).}. 
The authors furthermore propose to check for abnormal values. 
But abnormal values (i.e., spikes in the measured clock offset) only occur for simple delay attacks but not for incremental delay attacks (as to be shown in Section~\ref{sec:experiment}). 

\gls{ptp} version 2 includes an experimental security extension, Annex K~\cite{ptp}, which provides message integrity and replay protection. 
In Annex K, however, several flaws were discovered and it was never properly formalized~\cites{treytl_security_2009}{moreira_security_2015}. 
In any case, Annex K does not cover attacks in the time domain at all. 
Interestingly, the \gls{ptp} standard~\cite{ptp} mentions link asymmetry (although in a single sentence only) but does not cover delay attacks any further. 
We analyze asymmetric link delay attacks in Section~\ref{sec:asymmattack} and countermeasures against such attacks in Section~\ref{sec:asymmcounter}. 

\gls{nts}~\cites{ietf-ntp-network-time-security}{ietf-ntp-using-nts-for-ntp}{ietf-ntp-cms-for-nts-message} consist of a set of \gls{ietf} drafts that aim at providing authenticity and integrity for unicast clock synchronization protocols. 
Until now, \gls{nts} is only specified for \gls{ntp} and covers delay attacks briefly. 
A recently proposed security extension to \gls{ptp}~\cites{itkin_security_2016-1}{itkin_security_2016} aims to secure the \num{2}-step mode in \gls{ptp}. 
The authors mention delay attacks in the extended version of their paper but refer to related work when it comes to countermeasures. 

\citeauthor{ptpv3macsec} as well as \citeauthor{4659223} recommend MACsec to secure clock synchronization~\cites{ptpv3macsec}{4659223}. 
\citeauthor{mizrahi_time_2011} analyzes IPsec and MACsec as means to secure clock synchronization~\cite{mizrahi_time_2011}. 
Delay attacks are very briefly mentioned but the paper neither has details on how delay attacks can be conducted or mitigated nor how IPsec and MACsec relate to delay attacks. 
\citeauthor{treytl_securing_2010} discuss the usage of IPsec to protect \gls{ptp}~\cite{treytl_securing_2010} but do not mention delay attacks at all. 
\citeauthor{mizrahi_game_2012} states that encrypting clock synchronization traffic makes it more difficult to conduct delay attacks~\cite{mizrahi_game_2012}. 
We will show that this holds true to some extent for selective messages attacks only (Section~\ref{sec:selectiveattack}), but asymmetric link delay attacks (Section~\ref{sec:asymmattack}) are not obstructed by encryption at all. 

In~\cite{6645332}, \citeauthor{6645332} simulated delay attacks and propose a countermeasure based on hypothesis testing. 
\citeauthor{moreira_security_2015} discuss delay attacks briefly~\cite{moreira_security_2015}. 
\citeauthor{ullmann_delay_2009} examined delay attacks on \gls{ptp} and \gls{ntp}~\cite{ullmann_delay_2009} and propose what we call \textit{bound clock offset by limiting the \gls{rtd}} as mitigation. 
\citeauthor{lisova_protecting_2016} provide a good analysis of the consequences of delay attacks and discusses \gls{rtd} limits as countermeasure~\cite{lisova_protecting_2016}. 
Other countermeasures discussed, such as monitoring interarrival times, are unreasonable in a general setting because those times will be equally affected by a delay attack and cannot be used therefore as a countermeasure against delay attacks. 
\citeauthor{mizrahi_game_2012} also briefly discusses \gls{rtd} limits as a potential countermeasure against delay attack~\cite{mizrahi_game_2012}, and so do \citeauthor{securetime}~\cite{securetime}. 

\citeauthor{mizrahi_game_2012} proposes to use multiple paths between master and slave to mitigate delay attacks~\cite{mizrahi_game_2012}. 
The assumptions are quite strong, however, as all paths need to be entirely (i.e., also physically) independent and the adversary may only attack a minority of paths successfully. 
From a practical perspective it seems unrealistic to provide multiple low-latency paths with low delay variation that are entirely independent. 
It is implicitly assumed that the various networking components such as \glspl{nic}, routers, and switches are from distinct vendors and that paths are symmetric. 
Furthermore, the countermeasure does not scale well as it is more costly to establish new independent paths than to compromise any majority of paths. 

In recent work \citeauthor{8357814} developed a mathematical model that defines necessary and sufficient conditions for secure clock synchronization~\cite{8357814}. Focusing on conditions that must be met in wireless networks to detect \gls{mitm} attacks on clock synchronization, the paper builds on earlier work~\cite{securetime} to conclude that round-trip message exchange is a prerequisite for secure clock synchronization and that one-way clock synchronization can not be secured against delay attacks. However, \citeauthor{8357814} rely on uniquely identifiable clock synchronization messages in an isolated context and maximum round-trip delays, whereas our work analyses the much broader context of clock synchronization as part of real, encrypted traffic.  

Summarizing related work, research on secure clock synchronization often either excludes delay attacks or refers to related work for discussion on this issue. None of the related work discusses delay attacks on encrypted clock synchronization traffic in a realistic context, including detection of clock synchronization traffic.
Furthermore, related work does not recognize the ability of clock synchronization slaves to determine guaranteed bounds on their clock offset at a specific point in time, nor does it identify the fundamental limitation inherent to network-based clock synchronization: clock synchronization can either be high-precision or secure against delay attacks. 
 	
	\section{Selective Message Delay Attacks}
	\label{sec:selectiveattack}

To secure communications over untrusted networks the entire communication is commonly encrypted and authenticated, for instance with IPsec. 
For this reason, we tested delay attacks with IPsec in tunnel mode\footnote{We have tested selective message delay attacks successfully against commercially available systems like routers or protections switches that tunnel \gls{ptp} using security protocols different than IPsec, too. Methods and conclusions are identical to the ones presented for IPsec tunnels for all tested systems.}. %
In such scenario, the attacker has access only to encrypted traffic, which means that there is no information available about protocols, source and destination ports, nor the IP addresses of the real endpoints. 

In this section, we show that clock synchronization messages can be reliably identified in an encrypted traffic stream with reasonable effort. 
This identification of (encrypted) clock synchronization messages builds the foundation for selective message delay attacks. 
For this purpose, we aim to answer the following questions:
\begin{itemize}
\item Are there any (statistical) properties of \gls{ptp} traffic that can be used to identify \gls{ptp} messages within encrypted traffic?
\item Can \gls{ptp} traffic be modified such that delay attacks can be prevented or at least mitigated?
\item Can encryption schemes provide reasonable security against selective message delay attacks? 
\end{itemize}
To answer whether it is possible to identify both \gls{ptp} traffic in general and specific types of \gls{ptp} messages in encrypted traffic, we conduct statistical traffic analysis of \gls{ptp} traffic. 
From this analysis, we identify several properties of \gls{ptp} traffic that build the foundations for selective message delay attacks. 
We furthermore implement selective message delay attacks on actual devices to show the feasibility of our proof of concept in practice.

\subsection{PTP Traffic Analysis}
\label{sec:trafficanalysis}
In order to make a slave adhere to a false time, either the \sync or the \delayreq message need to be delayed by the adversary (as pointed out in Section~\ref{sec:delayattacks}). 
Depending on the delaying of either \sync or \delayreq, the slave's notion of time is going to be behind or ahead of the master's, respectively. 
When traffic is encrypted, traffic analysis is limited to a restricted set of features: time, packet length, source, and destination.  
In our specific case, each of these features can be used to identify the particular type of \gls{ptp} message in encrypted traffic. 
The statistical properties identified in \gls{ptp} in both phases, i.e., clock offset measurement and delay measurement, are closely related to timing, packet length, and packet direction. 

One \gls{ptp} clock synchronization cycle consists of a series of four messages as highlighted in Section~\ref{sec:back}. 
(1) A \sync message from the master to the slave. 
(2) Another message (\followup) from the master to the slave. 
(3) A message \delayreq in the reverse direction from the slave to the master, and (4) a message from the master to the slave (\delayresp). 
This series repeats at a fixed interval. 
Every two seconds we observe another \gls{ptp} message (\announce) from the master to the slave. 
The \announce message is used for the Best-Master-Clock algorithm, which we do not focus on in this paper. 

\begin{figure}
	\centering
	\includegraphics[width=0.9\linewidth]{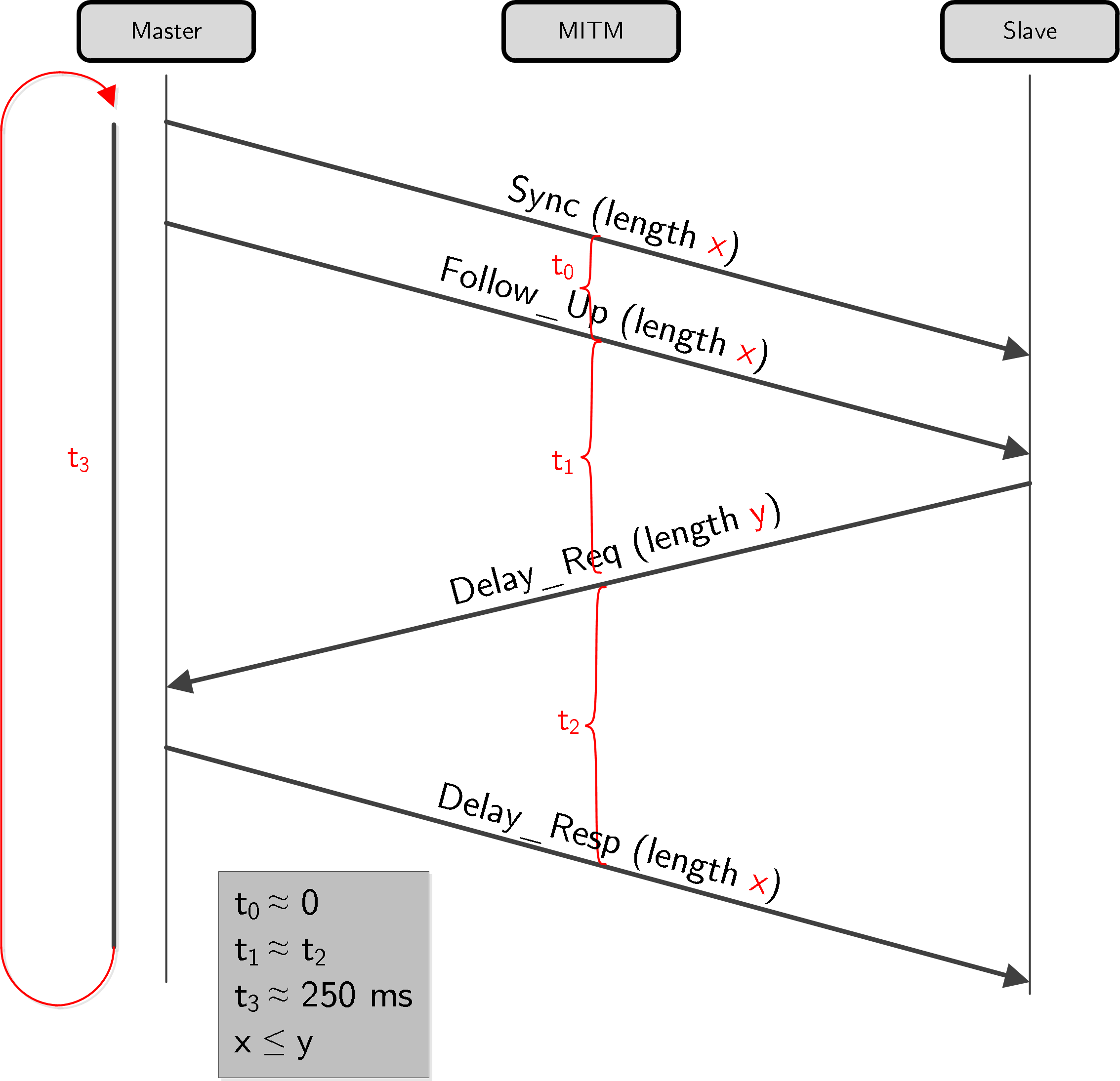}
	\caption{\gls{ptp} traffic patterns in timing, length, and direction.}
	\label{fig:ptp_analysis}
\end{figure}

Our traffic analysis revealed some specific properties of \gls{ptp} traffic. These properties are the length of packets, the timing, and the direction of messages. 
The lengths of the packets are appealing as they are highly deterministic and mostly constant. 
The reason for this is that the designers of \gls{ptp} wanted to avoid variation in the transmission delays because of differences in packet lengths. 
Fig.~\ref{fig:ptp_analysis} sketches the results of our \gls{ptp} traffic analysis. 
Firstly, all messages from the master to a slave are of the same length and the \delayreq message in the reverse direction is either of equal length or slightly longer, which means that the length of a \gls{ptp} message is related to its direction. 
The specific lengths of the messages depend on the underlying communication protocols and on which layer the messages are observed. 
In our setup, the lengths of (unencrypted) \gls{ptp} messages were \SI{86}{\byte} and \SI{96}{\byte} for \sync, \followup, and \delayresp and \delayreq respectively (and \SI{106}{\byte} for \announce messages). 
In encrypted traffic, additional information is added to the packet by the encryption scheme, increasing the packet's length. 
The packet lengths observed were \SI{138}{\byte} and \SI{154}{\byte} in a test with IPsec encryption. 
Other encryption schemes result in different lengths but the observed pattern persists. 
Messages of length \SI{154}{\byte} occur every \SI{2}{\second}, which corresponds to \announce messages. 
The remaining packets with a length of \SI{138}{\byte} occur every \SI{250}{\milli\second} and in sets of \num{4} (which corresponds to the \sync, \followup, and \delayreq and \delayresp messages). 
The length of encrypted \gls{ptp} messages are deterministic, as well, and therefore identifiable. 

Secondly, we observe that \gls{ptp} messages follow a specific timing pattern. 
The \followup message is sent from the master with minimal delay ($t_0$) after the \sync message; the \delayreq message is sent with a small delay ($t_1$) after the \followup message was received by the slave; and the \delayresp is also sent immediately after the \delayreq was received ($t_2$) by the master. 
From the adversary's point of view $t_1$ and $t_2$ are roughly similar, which is about the \gls{owd} between master and slave\footnotemark. 
Because of their periodicity, the observed timings are also visible in encrypted traffic, as we will see later in this section. 

\footnotetext{$t_1$ and $t_2$ are roughly similar if the attacker is positioned equidistant in terms of time between master and slave. If the attacker is closer to either master or slave, the difference between $t_1$ and $t_2$ increases --- visible by shifting the observation point to the left or to the right, respectively, in Fig.~\ref{fig:ptp_analysis}.}

Thirdly, the direction of the messages is fixed as long as the master and slave roles persist. 
And this pattern repeats periodically at a fixed interval ($t_3$) for as long as the clock synchronization service is running. 
This clock synchronization interval can be configured and was left to the default setting of \SI{250}{\milli\second} during our tests. 
Table~\ref{fig:analysis} summarizes the results of our traffic analysis. 
In Section~\ref{sec:matlab} we show how these properties can be used to identify \gls{ptp} traffic in a stream of encrypted traffic. 

\begin{table*}[!htb]
\small
\renewcommand{\arraystretch}{1.2}
\caption{Identified properties of \gls{ptp} traffic.}
\centering
\begin{threeparttable}
\label{fig:analysis}
\begin{tabular}{l c c c}
	\toprule
	\bfseries Message type & \multirow{2}{*}{\shortstack{\bfseries Direction\\ \bfseries(master-slave)}} & \bfseries Length                  & \bfseries Timing                        \\
						   & & & \\ \midrule\midrule
	\sync                  & $\rightarrow$       & \SI{86}{\byte} / \SI{138}{\byte}  & $t_3$                                   \\
	\followup              & $\rightarrow$       & \SI{86}{\byte} / \SI{138}{\byte}  & $t_0$                                   \\
	\delayreq              & $\leftarrow$        & \SI{96}{\byte} / \SI{138}{\byte}  & $t_1 \approx t_2 \approx$ one-way delay \\
	\delayresp             & $\rightarrow$       & \SI{86}{\byte} / \SI{138}{\byte}  & $t_2 \approx t_1 \approx$ one-way delay \\
	\announce              & $\rightarrow$       & \SI{106}{\byte} / \SI{154}{\byte} & fixed interval (ignored)                          \\ \bottomrule
\end{tabular}
\begin{tablenotes}
\small
\item $t_3$ with regard to the last \delayrespsmall message, i.e., the clock synchronization interval.
\end{tablenotes}
\end{threeparttable}
\end{table*}

\subsection{Identification of PTP messages in Encrypted Traffic}
\label{sec:matlab}
In order to conduct a selective message delay attack on an encrypted traffic stream, we first need to identify the specific PTP messages within the encrypted traffic without prior knowledge of the communications within that stream. 
To this end, we setup a proof of concept to verify that specific PTP messages can be identified within encrypted traffic. 
In a real-world scenario, the specific setup will always be different, and attackers may not have the plaintext communication available to figure out the specific parameters of the setup (i.e., the specific packet sizes, the PTP session interval, etc). 
Also, packet lengths differ in encrypted traffic (as there is additional data added by the encryption scheme) but the basic properties (frequency, direction, timing and the relation of the lengths) persist.

Using the results of our \gls{ptp} traffic analysis, we wanted to know whether identifying \gls{ptp} messages in encrypted traffic is possible. 
For this purpose, we conducted some experiments in which \gls{ptp} communications were simulated.
The communication under test selects random timings $t_0$, $t_1$, $t_2$, $t_3$, and random lengths $x$ and $y$.  
The simulation takes two parameters, the noise level and the time to observe the traffic. 
Noise is added to the simulated \gls{ptp} communication, which is the probability to obtain a non-\gls{ptp} packet of random length every millisecond. 
Eventually, the script tries to estimate $t_0$, $t_1$, $t_2$, $t_3$, and the lengths $x$ and $y$ just by observing the communication.

This proof of concept relies on four assumptions:
\begin{enumerate}
\item Time is discretized with a sampling time of \SI{1}{\milli\second}.
\item There is only one packet per time bin.
\item The pattern repeats all over the observation period.
\item $t_0$, $t_1$, $t_2$, and $t_3$ are constant from the perspective of the sampling time.
\end{enumerate}

Such assumptions draw some limitations for simulating communications, but they are assumable since experiments are intended to be proofs of concept and not fully-fledged implementations. 
Real-world communications comprise additional complexities that might defy the proof of concept detector, but such situation can usually be faced by refining the detector (for example, by using recurrence analysis, granger causality, or markov models). 
Our goal is to show that the detection of the pursued time parameters and lengths is theoretically possible and feasible by the application of methods based on statistics. 

Our simulations show that packet lengths, directions, and timing are sufficient to separate \gls{ptp} from other traffic and even to identify the particular type of \gls{ptp} message so that the selective message delay attack can be conducted. 
If the noise level is increased, the observation time needs to be increased as well (as expected).
Under our simulation conditions, at a noise level of \num{99.9}\%, we need to observe the communications for \num{1000} seconds to reliably determine the particular times and lengths.  

A challenging scenario for the detection would be the occurrence of periodic signals with similar properties (timing, lengths, and direction) within the same flow. 
We argue that the properties of \gls{ptp} are very particular and the chances to encounter communications with similar properties in the same traffic are very low. 
In the light of our experimental results, we conclude that \gls{ptp} traffic can be identified with high probability within encrypted network traffic. 
For this reason, selective message delay attacks can be conducted on PTP even when it is (supposedly) secured with state of the art network encryption schemes such as IPsec.

\subsection{Experimental Results}
\label{sec:experiment}
\begin{figure}
	\centering
	\includegraphics[width=0.9\linewidth]{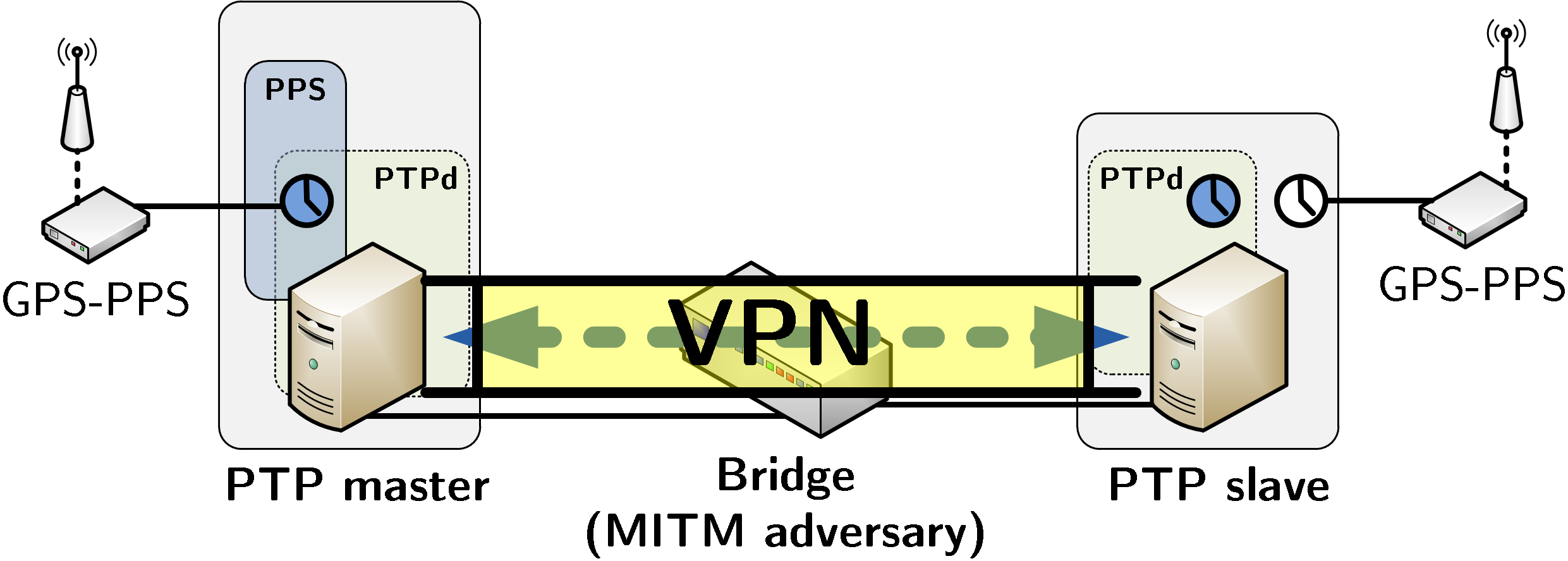}
	\caption{Experimental setup used to examine the effect of delay attacks on \gls{ptp} over an untrusted network.}
	\label{fig:versuch}
\end{figure}

Based on the statistical properties we identified in the traffic analysis (Section~\ref{sec:trafficanalysis}) and the theoretical confirmation that these properties can be used to identify \gls{ptp} messages within encrypted traffic (Section~\ref{sec:matlab}), we implemented a \gls{ptp} traffic detection and \gls{ptp} message type identification on a real clock synchronization system. 
We furthermore implemented selective message delay attacks on actual devices to show the feasibility of our proof of concept in a practical setting. 
Fig.~\ref{fig:versuch} depicts the experimental setup we used to evaluate the feasibility of our proof of concept and to examine the effect of delay attacks on real \gls{ptp} systems over untrusted networks. 
We used three Linux systems: one that runs as \gls{ptp} master, another as \gls{ptp} slave, and the third acts as network bridge. 
Master and slave were connected through an IPsec tunnel such that the bridge could only observe encrypted traffic. 
\gls{ptp} master and slave both run PTPd version 2.3.1. 
Master and slave receive a \gls{gps} \gls{pps} signal but only the master clock is synchronized to it. 
With this setup, we can synchronize the master clock $\pm$\SI{10}{\micro\second} to \gls{utc}, which is not overly precise but enough to highlight the effect of selective message delay attacks. 
The slave clock is synchronized to the master via \gls{ptp}. 
To compare the slave clock to the \gls{pps} signal, the “ppstest” tool\footnote{\url{https://github.com/redlab-i/pps-tools}} was used. 

Both systems are connected through an Ethernet bridge. 
On the bridge, we implemented a \gls{mitm} application that can delay specific packets. 
The bridge is implemented with libnetfilter\_queue\footnote{\url{https://netfilter.org/projects/libnetfilter_queue/}} so that packets are not only available in kernelspace but also in userspace, which facilitates easier classification and attack implementation. 
Since traffic is encrypted, a delay attack in the value domain does not work because the timestamps within the packets cannot be modified. 
Instead, the attack only works in the time domain. 
In general, the adversarial application on the bridge aims to identify \gls{ptp} messages and delay specific messages as soon as the selective message delay attack is started. 
We expect that the slave clock is then desynchronized from the master clock after a short time (due to an averaging algorithm employed in PTPd). 

\begin{figure}
	\centering
	\includegraphics[width=0.9\linewidth]{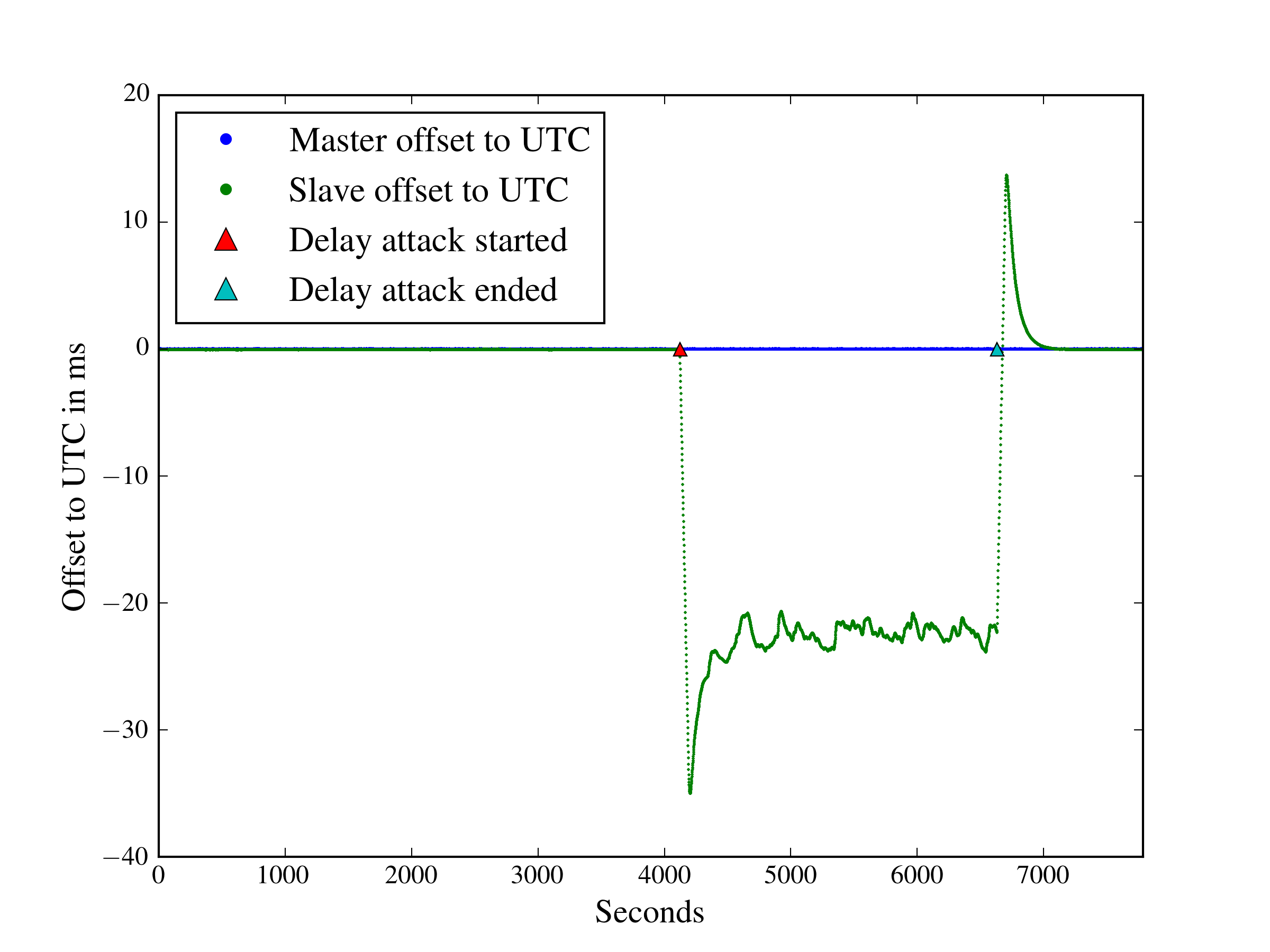}
	\caption{Offset to UTC during an selective \syncsmall message delay attack.}
	\label{fig:attack20}
\end{figure}

\subsubsection{Experiment 1: Delayed \sync messages}
In the first experiment, we programmed our malicious application to delay all \sync (and \followup) messages by~\SI{50}{\milli\second}. 
We chose such large delay to stress that during such an attack the clock synchronization cannot be considered high-precision any more. 
Fig.~\ref{fig:attack20} shows the results, i.e., the offset of the master and of the slave clock to \gls{utc} during normal operation and during the attack. 
The master clock is quite stable throughout the run of the experiment, as expected. 
The slave clock spikes\footnote{Presumably, the overshooting in those spikes at the begin and at the end of the attack is caused by the specific control algorithm implementation in PTPd.}~shortly after the attack is started (at time \num{4122}) and ended (at time \num{6632}) and settles around~\SI{-25}{\milli\second} to \gls{utc} after a couple of seconds throughout the attack. 
The master clock is not affected at all by the selective \sync-message delay attack, as expected. 
The slave clock is affected, however, since the clock synchronization messages of the master are delayed maliciously. 
The delay attack operates as intended since the slave clock is around~\SI{25}{\milli\second} behind the master's during the attack. 

\subsubsection{Experiment 2: Delayed \delayreq messages}
\begin{figure}
	\centering
	\includegraphics[width=0.9\linewidth]{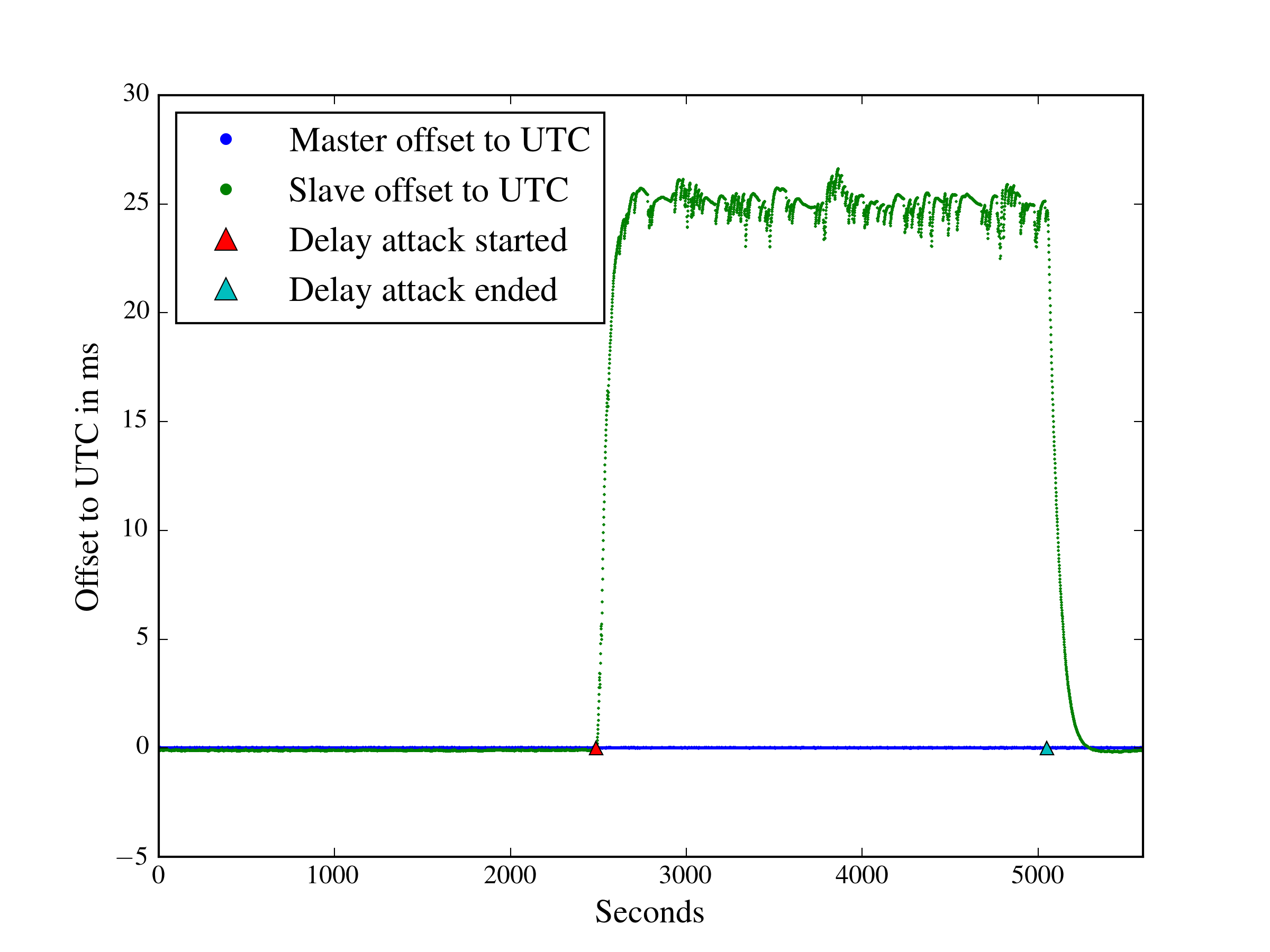}
	\caption{Offet to UTC during a selective \delayreqsmall message delay attack.}
	\label{fig:attack23}
\end{figure}

The same setup was used in a second experiment. 
This time, \delayreq messages were delayed \SI{50}{\milli\second} by our adversarial application on the bridge (instead of \sync and \followup messages). 
For this reason the slave clock is (roughly \SI{25}{\milli\second}) ahead of the master's clock during the attack as shown in Fig.\vref{fig:attack23}. 

\subsubsection{Experiment 3: Incremental delay attack}
\begin{figure}
	\centering
	\includegraphics[width=0.9\linewidth]{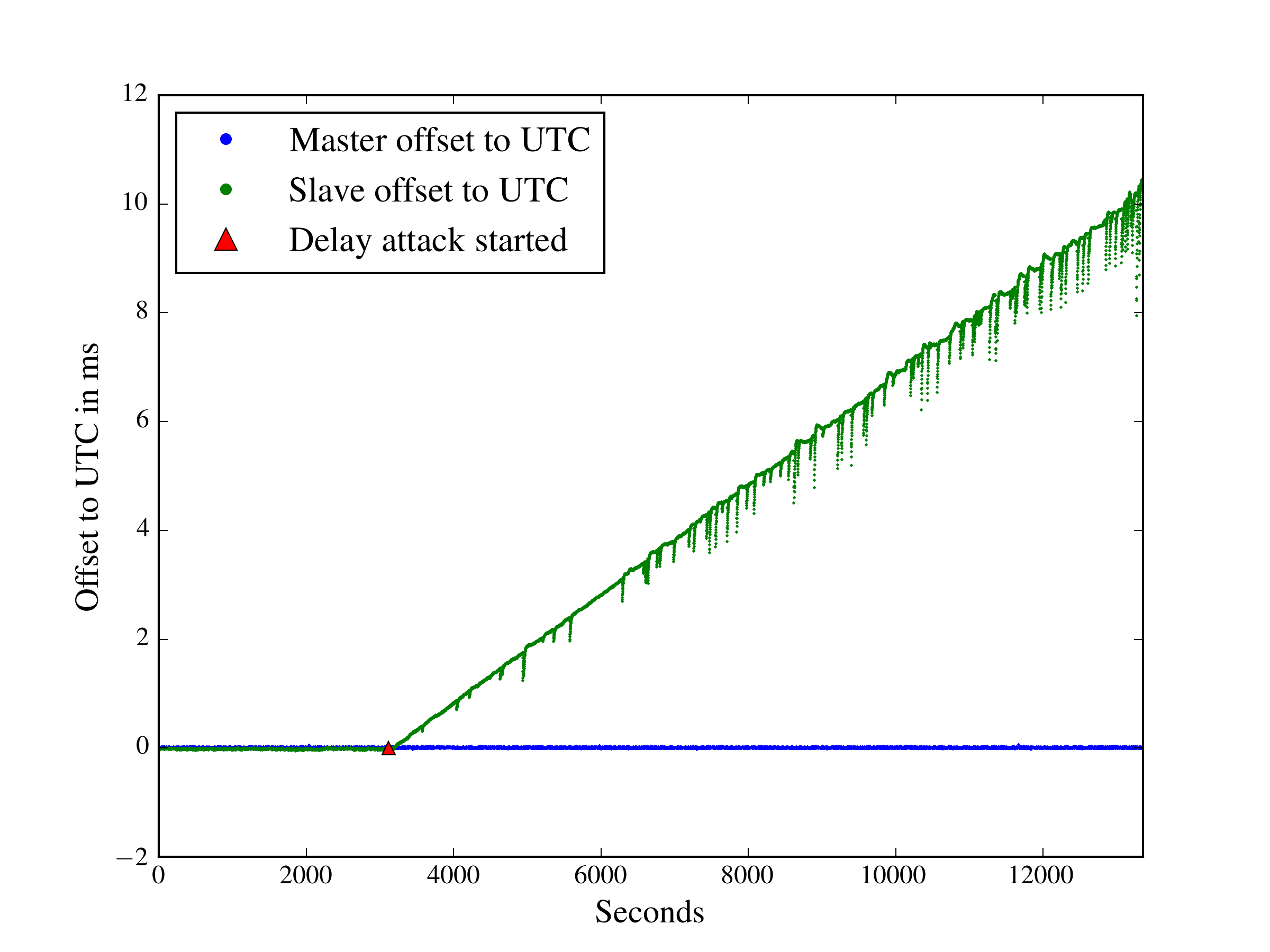}
	\caption{Offset to UTC during a incremental selective \delayreqsmall message delay attack.}
	\label{fig:attack22}
\end{figure}

One may argue that the spikes in clock offset at the start and end of the attacks may raise a suspicion in security-critical environments. 
For this reason, we also implemented selective message delay attacks that incrementally add malicious delay. 
In that case (shown in Fig.~\ref{fig:attack22}), the attacker does not apply the full malicious delay from the moment the attack is started but instead incrementally increases the malicious delay with each message. 
In this way, there is no more spike in the slave's time offset when the attack starts. 
At time \num{3119} the incremental delay attack was started as \delayreq messages were (increasingly) delayed by \num{1} ppm. 

This incremental delay was deliberately chosen small (with \num{1} ppm) to highlight that such small delay is indistinguishable from delay variation while still having a significant effect on the clock's precision. 
Despite the small incremental delay, the slave clock is off more than~\SI{7}{\milli\second} to \gls{utc} after two hours, which is completely unacceptable for most time critical systems that rely on a precise notion of time. 
The slave cannot notice the drift of its clock relative to the master's because of the incremental delay attack and is therefore convinced to be perfectly synchronized. 

\subsubsection{Discussion}
While the first two attacks might be detectable due to the spikes in the clock offset at the start and end of the attacks, the incremental attack cannot be detected easily. 
Moreover, we argue later in Section~\ref{sec:asymmcounter} that such incremental delay attack cannot be prevented at all (only be mitigated to some extent under specific circumstances). 
Although the attacker can neither see the packets' contents, nor the real endpoints, nor ports, selective message delay attacks can be conducted successfully. 
This indicates that encryption alone cannot prevent selective message delay attacks on clock synchronization.

	\section{Countermeasures Against Selective Message Delay Attacks}
	\label{sec:selectivecounter}

In order to secure \gls{ptp} from selective message delay attacks, encryption of the communication is not sufficient. 
But there exist two options to counter selective message delay attacks: (1) prevent traffic analysis, and (2) mitigate the actual attack. 
Both options will be discussed in this section.

\subsection{Traffic Analysis Mitigation}
Direction, timing, and packet length are sufficient to reliably identify \gls{ptp} traffic and the specific types of \gls{ptp} messages.  
In traffic analysis mitigation, the goal is to disturb traffic analysis. 
For this purpose, we need to make sure that the observable features entail no information that can be used to conduct an attack. 
Those features are packet length, time, and direction. 

While packet lengths are usually highly deterministic and constant, they can be hidden by padding to a fixed length. 
Such padding can even be implemented without changing the clock synchronization protocol. 
\gls{esp} mode in IPsec, for example, supports payload padding up to \num{255} padding bytes~\cite{rfc4303}, and the Extension Header could be used in IPv6~\cite{rfc2460}. 
Alternatively or additionally \gls{tfc}~\cite{tfc} could be employed to ensure that all \gls{ptp} messages have the same length. 
Alternatively, random lengths could be used, but this would increase variation of transmission delays (in addition to increasing bandwidth requirements), which is detrimental to the goal of achieving high-precision clock synchronization. 

Nevertheless, padding alone is not sufficient for traffic analysis mitigation since information about the time and the direction are still sufficient to identify \gls{ptp} packets (and therefore to conduct selective message delay attacks). 
The next feature that needs to be changed is the timing. 
Because of the periodicity of \gls{ptp} messages discovered in our \gls{ptp} traffic analysis, the specific messages can be identified reliably even in encrypted traffic.
Changing the timing, however, can only be done from within the clock synchronization protocol and not by external mechanisms. 
The offset measurement and delay measurement could be separated, the offset correction not executed periodically but in random intervals, and there could be a random interval as well between \sync and \followup messages. 
In this way, the timing properties of \gls{ptp} could hardly be used anymore to identify \gls{ptp} packets reliably under the assumption that sufficient cover traffic exists with similar timing properties. 
The major downside of this method is that it depends on the continuous existence of suitable cover traffic over the entire path from master to slave, nevertheless. 
This prerequisite of suitable cover traffic shifts the discussion to the well-researched area of traffic obfuscation in order to protect \gls{ptp} from selective message delay attacks (even when traffic is encrypted). 
Traffic obfuscation is known to be very complex and its security highly depends on the specific threat model\footnote{A special case here is that other periodic signals (that show similar properties to those identified in \gls{ptp}) could be used to mitigate traffic analysis.}. 

The last feature that is used to prepare selective message delay attacks is the direction of messages. 
As we have seen from the traffic analysis (Section~\ref{sec:trafficanalysis}), the packets' directions are highly deterministic in \gls{ptp}. 
There is no straight-forward way to remove this feature.

\subsection{Mitigation of Delay Attacks}
In this section, we highlight two techniques to mitigate delay attacks. 
The first technique relies on knowledge of the underlying communication network and requires changes to the clock synchronization protocol. 
The second technique builds upon the replay protection of the encryption scheme. 
It needs to be stressed, however, that both techniques are mitigation only and cannot prevent the attacks. 
Table~\ref{fig:selectivecounter} summarizes the results. 

\begin{table*}[!htb]
\renewcommand{\arraystretch}{1.2}
\caption{List of countermeasures against selective message delay attacks.}
\centering
\label{fig:selectivecounter}
\begin{tabular}{l l l c}
	\toprule
	\multirow{2}{*}{\bfseries Countermeasure}          & \multicolumn{1}{c}{\multirow{2}{*}{\bfseries Benefit}}                              &
	\multicolumn{1}{c}{\multirow{2}{*}{\bfseries Drawbacks}}                              &
	\multirow{2}{*}{\shortstack{\bfseries Prevents\\ \bfseries attack}} \\ 
	& & & \\ \midrule\midrule
	Random send and reply times         & Confuses traffic analysis based on timings & \multirow{2}{4.7cm}{Requires protocol changes and depends on cover traffic.}                                  & No                        \\
	& & & \\
	Equal (or variable) message lengths & Confuses traffic analysis based on lengths & \multirow{2}{4.7cm}{Variable lenghts reduce precision and depends on cover traffic.}                                & No                        \\
	& & & \\
	Strict replay protection            & Limits max. impact of the attack           & Not feasible in all scenarios                                 & No                        \\
	Limiting \glspl{owd}             & Limits max. impact of the attack           & \multirow{2}{4.7cm}{Requires knowledge of the underlying communication network.}                                & No\\
	& & &\\
	\bottomrule
\end{tabular}
\end{table*}

As the name suggests, the replay protection of a network security protocol is supposed to prevent or mitigate replay attacks (and has not been specifically designed to prevent delay attacks). 
At the same time, however, the replay protection also limits the maximum impact of selective message delay attacks since packets cannot be delayed arbitrarily. 
The impact of selective message delay attacks may be limited by maxing the encryption scheme's replay protection and assuring that a sufficient number of packets per clock synchronization interval are sent through the network as cover traffic. 
Replay protection usually works as follows: a sequence number is added to the packet, and the receiver accepts packets with strictly increasing sequence numbers only or with sequence numbers from a certain window to allow packets to overtake other packets in the network. 
While not specifically designed for this purpose, such replay protection also limits the impact of delay attacks since packets will not be accepted by the receiver if too many other packets have been received meanwhile. 
For this reason, the maximum impact of the selective packet delay attack is directly related to the replay protection and to the number of packets per clock synchronization interval at the network location the attacker has access to. 

Therefore, replay protection should be configured strictly (when possible) such that overtaking of packets is not allowed at all. 
Packets may still be delayed maliciously, however, until the subsequent packets arrive. 
Also, the attacker may drop or delay the subsequent packets as well in order to increase the malicious delay for the \gls{ptp} packets. 
If there are no additional security measures in place such delay or drop of packets will not raise any suspicion. 
In any case, the strict replay protection in conjunction with sufficient packets per clock synchronization interval may limit the impact of selective message delays. 
It needs to be stressed that this mitigation depends on additional cover traffic within the entire network path from master to slave (and vice versa) which may not be under the defender's control. 
Furthermore, strict replay protection is not feasible in all scenarios because reducing packet loss rate may be more important in some scenarios than stricter replay protection. 

Countermeasures asymmetric link delay attacks (i.e., limiting \glspl{owd} - to be introduced in Section~\ref{sec:asymmcounter}) are also applicable as countermeasures against selective message delay attacks (but countermeasures against selective message delay attacks are not applicable to asymmetric link delay attacks). 
 	
	\section{Asymmetric Link Delay Attacks}
	\label{sec:asymmattack}

In Section~\ref{sec:selectiveattack} we introduced selective packet delay attacks, in which the attacker aims to identify \gls{ptp} messages (in an encrypted traffic stream) in order to delay specific \gls{ptp} packets selectively. 
In this section, we present an additional delay attack, which we denote in the following as \textit{asymmetric link delay attack}. 
In such asymmetric link delay attack, the attacker delays all packets in one direction of the link but not in the other (e.g., all packets from the master to the slave are delayed maliciously but those from slave to master are not). 

\begin{figure}
	\centering
	\includegraphics[width=0.9\linewidth]{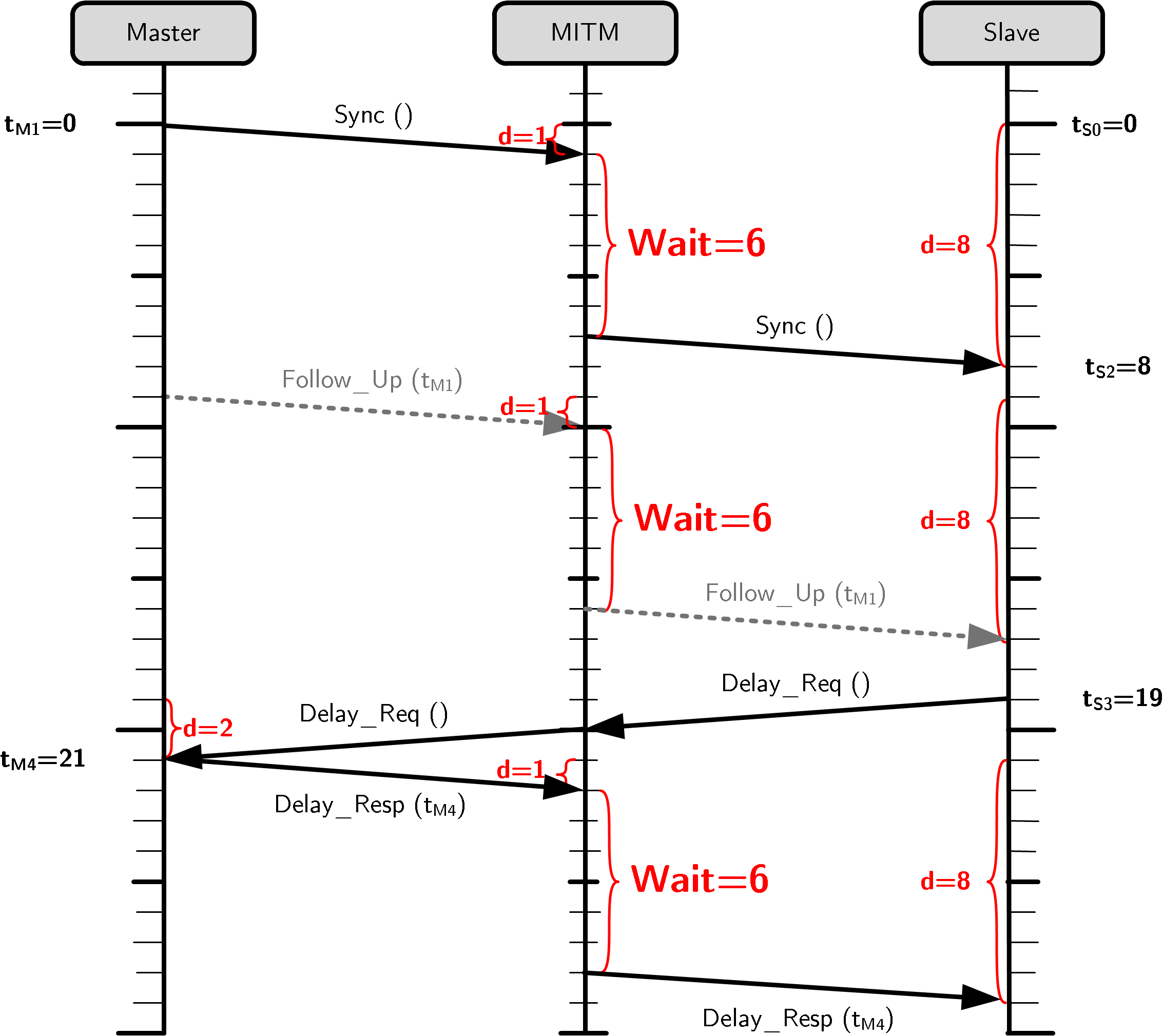}
	\caption{Asymmetric link delay attack where the master$\rightarrow$slave link is delayed.}
	\label{fig:link_asymmetry1}
\end{figure}

\begin{figure}
	\centering
	\includegraphics[width=0.9\linewidth]{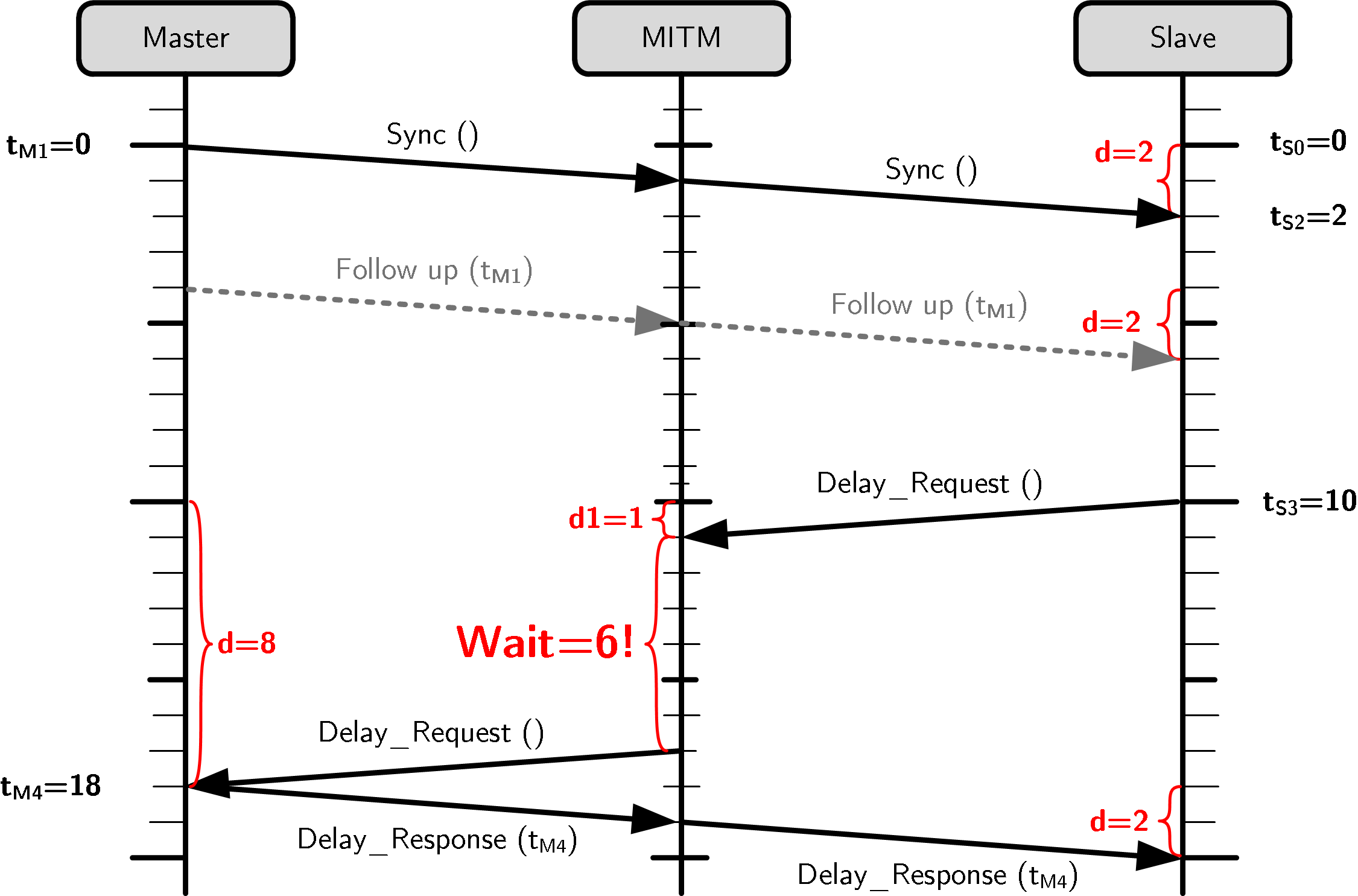}
	\caption{Asymmetric link delay attack where the slave$\rightarrow$master link is delayed.}
	\label{fig:ptp_attack3}
\end{figure}

In order to analyze how delaying \textbf{all} packets in one direction affects clock synchronization we use the example discussed in Section~\ref{sec:delayattacks}. 
This time, however, we delay \textbf{all} packets sent by the master to the slave without delaying the messages sent by the slave to the master as illustrated in Fig.~\ref{fig:ptp_attack2}. 
The slave eventually calculates the offset according to Eq.~\eqref{eq:offset} as \num{3} (again we assume for simplicity reasons that clocks are neither desynchronized nor drifting). 
The miscalculated clock offset (\num{3}) is half of the delay asymmetry while the real clock offset is \num{0}. 
For this reason, the slave sets its clock backwards by \num{3} time units. 
If the attacker conducts the attack in reverse direction (i.e., the packets from slave to master are maliciously delayed but the packets from master to slave are not, see Fig.~\ref{fig:ptp_attack3}), the slave miscalculates the offset as \num{-3} (according to Eq.~\eqref{eq:offset}) when it is actually \num{0}. 
For this reason, the slave will set its clock ahead by \num{3} time units. 

An attacker who conducts an asymmetric link delay attack by delaying all packets in one direction can, therefore, manipulate the clock offset by half of the asymmetric delay. 
The attacker can also influence the sign of the malicious offset correction by choosing which direction of packets are delayed maliciously. 
Link asymmetry being closely related to clock offset correction, two questions arise: (1) what is the exact relation between clock offset correction and link asymmetry, and (2) can asymmetric link delay attacks be prevented?
These two questions will be examined in the remainder of this section. 

When analyzing clock synchronization messages in detail, the first message is sent at $t_{M1}$ by the master and received at $t_{S2}$ by the slave. 
In the case of a hypothetical zero-delay link, the difference $t_{S2} - t_{M1}$ would represent the exact offset of the slave clock relative to the master. 
As pointed out in Section~\ref{sec:back}, clock synchronization messages in real systems experience various (constant and stochastic) delays along their path from master to slave and vice-versa. 
The receiving time $t_{S2}$ as well as the slave-computed clock offset incorporate the sum of all of those delays. 

In practice, transmission and propagation delays account for main part of the clock synchronization message end-to-end delay. 
This is why clock synchronization protocols comprise delay measurement methods to infer on applicable delays in order to compensate for them. 
On top of these measurements, high-precision clock synchronization protocols such as \gls{ptp} propose dedicated functionality in intermediate devices (so-called transparent clocks in routers and switches) to compensate for queuing and processing delays within intermediate systems, even though those delays typically amount to a minor part of the overall delay. 

\begin{figure}
	\centering
	\includegraphics[width=0.9\linewidth]{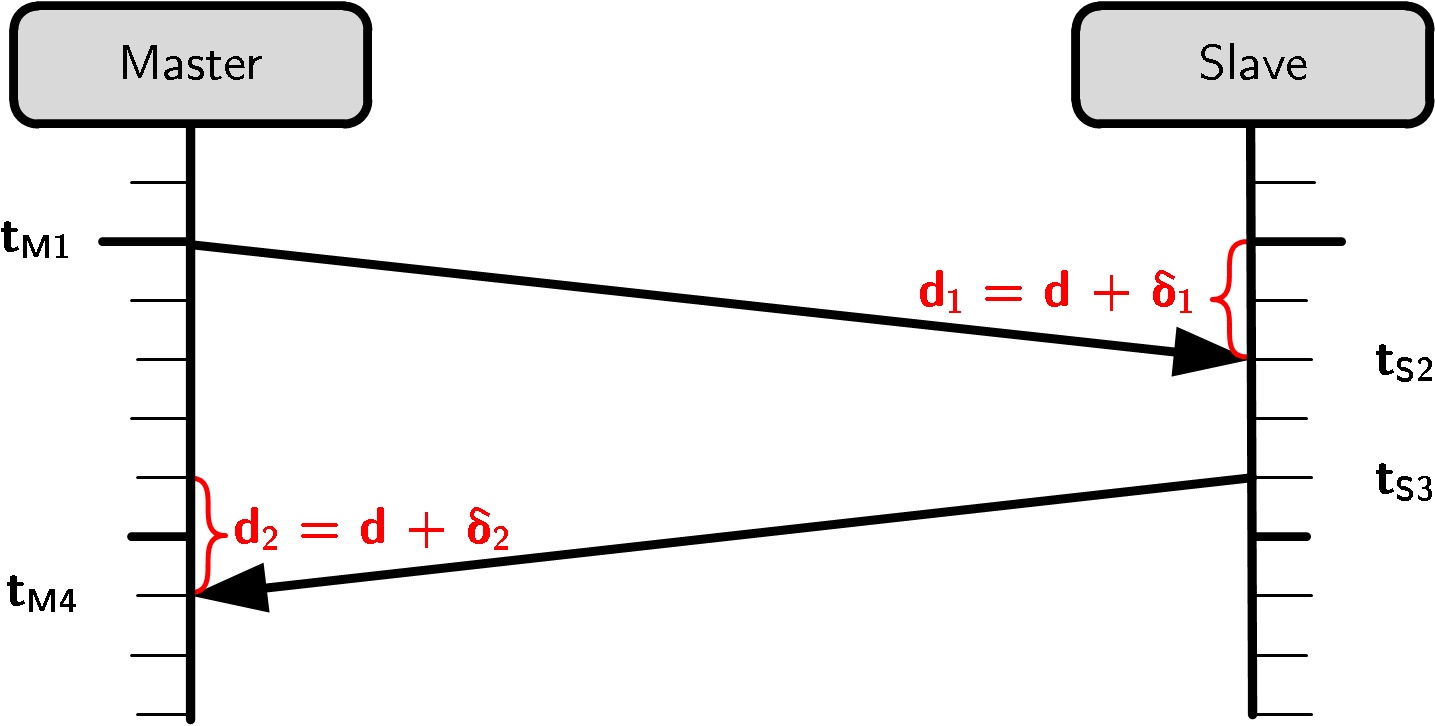}
	\caption{One-way delays in clock synchronization.}
	\label{fig:asymmetry}
\end{figure}

To facilitate the compensation of delays, two assumptions are required: (1) the relative clock drift within one measurement interval is negligible, and (2) the \glspl{owd} are symmetric, i.e., the sum of delays from master to slave equals the sum of the delays from slave to master. 
Fig.~\ref{fig:asymmetry} shows the \glspl{owd} $d_1$ and $d_2$ in the general case. 
The offset of the slave at time $t_{S2}$ is $offset(t_{S2}) = t_{S2} - t_{M1} - d_1$ and at time $t_{M4}$ is $offset(t_{M4}) = -t_{M4} + t_{S2} + d_2$. 
The offset is always calculated from the perspective of the slave so that the master inverts its offset calculation (as the offset of the slave clock relative to the master is the inverse of the offset of the master clock relative to the slave). 
The fact that always the slave's offset is calculated can be exploited by an attacker that maliciously alters delay asymmetry, even if the attacker does neither know the content of a message (because it is encrypted) nor its type (because traffic is perfectly obfuscated). The direction alone is sufficient to manipulate the measured clock offset. 

The assumptions mentioned above are essential to compensate the delays as the \glspl{owd} are approximated as half of the measured \gls{rtd}. 
If those assumptions do not hold true, the system of two equations and four unknown variables in Eq.~\ref{eq:offset_real} cannot be solved. 
The first assumption (relative clock drift is negligible during one measurement interval) implies that the clock offset at time $t_{S2}$ should be almost identical to the offset at time $t_{M4}$. 
For this reason, one variable is eliminated as $offset = offset(t_{S2}) = offset(t_{M4})$. 
The second assumption (symmetric delays) helps to eliminate another variable as $d = d_1 = d_2$, where $d_1$ is the delay from master to slave and $d_2$ is the delay from slave to master as shown in Fig.~\ref{fig:asymmetry}. This way, we end up having two equations with two variables that can be solved. 

In an asymmetric link delay attack an attacker exploits the assumption on symmetric delays. 
Assuming that there is a common part $d$ in \glspl{owd} and additional two distinct delays $\delta_1$ and $\delta_2$ for both directions (as shown in Fig.~\ref{fig:asymmetry}), the two offset equations are as follows:
\begin{equation}
\label{eq:offset_real}
\begin{split}
offset(t_{S2}) = t_{S2} - t_{M1} - d - \delta_1\\
offset(t_{M4}) = -t_{M4} + t_{S3} + d + \delta_2
\end{split}
\end{equation}
such that the offset can be calculated as
\begin{equation}
offset = \frac{t_{S2} - t_{M1} - t_{M4} + t_{S3} - \delta_1 + \delta_2}{2}
\end{equation}
as long as $offset(t_{S2}) \approx offset(t_{M4})$ holds true. 
The symmetric delay component $d$ is completely eliminated from the equation but $\delta_1$ and $\delta_2$ remain. 
If the delay is symmetric, then they cancel each other (as $\delta_1 = \delta_2$) so that the offset can be calculated precisely. 
But if the delays are not symmetric, offset calculation will be off by $\frac{\delta_2 - \delta_1}{2}$, which is in the interval $\left [\frac{-\delta_1}{2}, \frac{\delta_2}{2} \right]$ given that an attacker will eventually minimize either $\delta_1$ and keep $\delta_2$ close to zero or keep $\delta_1$ close to zero and maximize $\delta_2$. 
 	
	\section{Countermeasures Against Asymmetric Link Delay Attacks}
	\label{sec:asymmcounter}

In this section, we propose a method that facilitates defining maximum guaranteed bounds for the clock offset of the slave relative to the master. 
For the following discussion we assume that master and slave clocks are synchronous at time $t_{S0} = T_{M1} = 0$ and that there is no clock drift during one clock synchronization interval. 
All \gls{ptp} messages are cryptographically signed and encrypted, so the \gls{mitm} can not modify timestamp values within these messages. 
The \followup message is omitted, and we assume that the slave immediately sends the \delayreq after reception of the \sync message, i.e., $t_{S3} = t_{S2}$, for sake of simplicity. 
However, it is important to stress that neither the initial clock synchronization nor the immediate sending of \delayreq are a prerequisite for the proposed method. 
This is essential for avoiding \delayreq message collisions from multiple slaves following \sync multicasts by the master, which could cause link-layer collisions in the network. 
However, the (in)equations below consider already separate timestamps  $t_{S2}$ and $t_{S3}$, so slaves can use random delays after receiving \sync to avoid potential collisions. 

\begin{figure}
	\centering
	\includegraphics[width=0.9\linewidth]{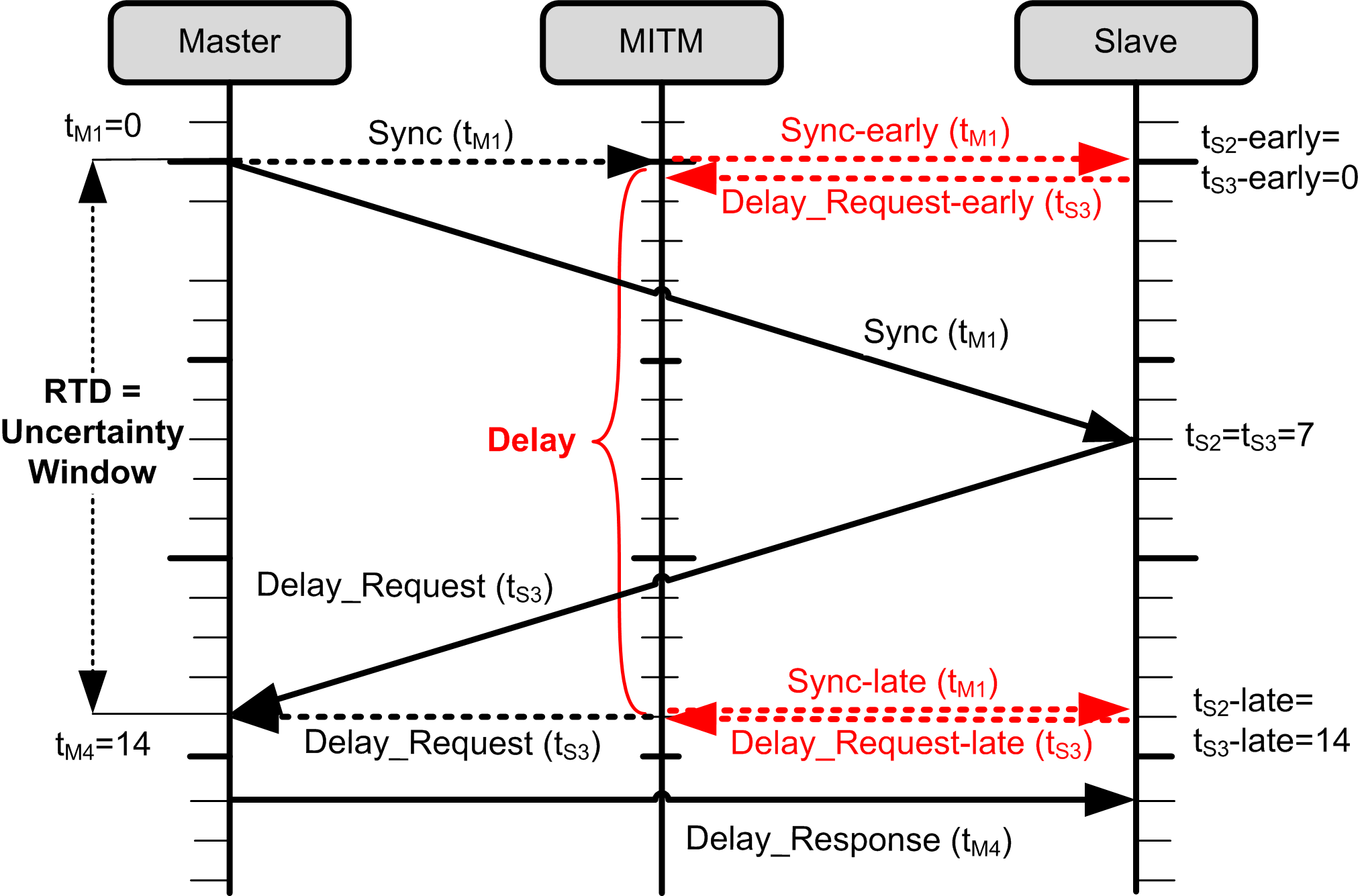}
	\caption{Theoretical (worst case) offset uncertainty bound calculation.}
	\label{fig:link_constraints-theoretical}
\end{figure}

In Fig.~\ref{fig:link_constraints-theoretical}, the master sends the \sync message at time $t_{M1} = 0$ and receives the slave's \delayreq at time $t_{M4} = 14$. 
The slave receives the master's timestamp $t_{M4} = 14$ in the \delayresp message and computes the \gls{rtd} as \num{14} (according to Eq.~\ref{eq:rtd}). 
The default assumption of \gls{ptp} is that communication paths are symmetrical and, therefore, the timestamps $t_{S2}=t_{S3}$ must be mapped to the center of the master's interval, i.e. $t_{S2}=t_{S3}=7$ as shown in Fig.~\ref{fig:link_constraints-theoretical}. 

Unless there is specific information available on physical delays for the forward and reverse communication path, it must be assumed that the delay of one or both of these paths can be (close to) zero. 
This gives an adversary the opportunity to arbitrarily delay the master's \sync message on the forward path or the slave's \delayreq message on the reverse path within the given window of \num{14} time units (as shown in Fig.~\ref{fig:link_constraints-theoretical}). 
The \gls{mitm} adversary can forward the master's \sync message without additional delay (\synce) and delay the slave's \delayreqe message by \num{14} time units, resulting in slave timestamps $t_{S2}{\text -}early=t_{S3}{\text -}early=0$. 
Alternatively, the adversary can delay the master's \sync message (\syncl) and forward the slave's \delayreql message without additional delay, resulting in slave timestamps $t_{S2}{\text -}late=t_{S3}{\text -}late=14$. 
Therefore, depending on which scenario the adversary \gls{mitm} adopts, it can shift the slave's offset within $\interval{-7}{7}$ time units (according to Eq.~\eqref{eq:offset}). 
Whenever the slave depends on maximum guaranteed bounds of its clock offset, this uncertainty must be considered.

\subsection*{Bound Clock Offset With Knowledge on OWDs}
\label{sec:rtdbounds}
In order to reduce the uncertainty and guarantee bounds on the clock offset, we present a method that builds upon knowledge on physical parameters and constraints of the clock synchronization's communication path. %
In this way, the attacker's ability to conduct delay attacks is reduced, and the slave is supported in determining stricter guaranteed bounds for its clock offset. 
It is worth noting that the method can only mitigate delay attacks but not prevent them entirely. 

We assume that the communication path is asymmetric and its minimum \gls{owd} is known for both directions. 
We denote the minimum \gls{owd} from master $\rightarrow$ slave as $d_{min}^{MS}$ and the minimum \gls{owd} for slave $\rightarrow$ master as $d_{min}^{SM}$. 
The exact measurement of \glspl{owd} depends on precisely synchronized clocks, which is why in real-world scenarios $d_{min}$ may be approximated using topology and physical parameters like propagation-, transmission- and processing delays of the network path's links and components. 
A mandatory precondition for guaranteed clock offset bounds is that the approximated \gls{owd} must be less or equal to the minimum real packet delay on the path. 
Conservative approximations on $d_{min}^{MS}$ and $d_{min}^{SM}$, i.e., lower minimum \gls{owd} values are detrimental to the offset bounds but are essential to guarantee the bounds in adversarial settings. 

\begin{figure}
	\centering
	\includegraphics[width=0.9\linewidth]{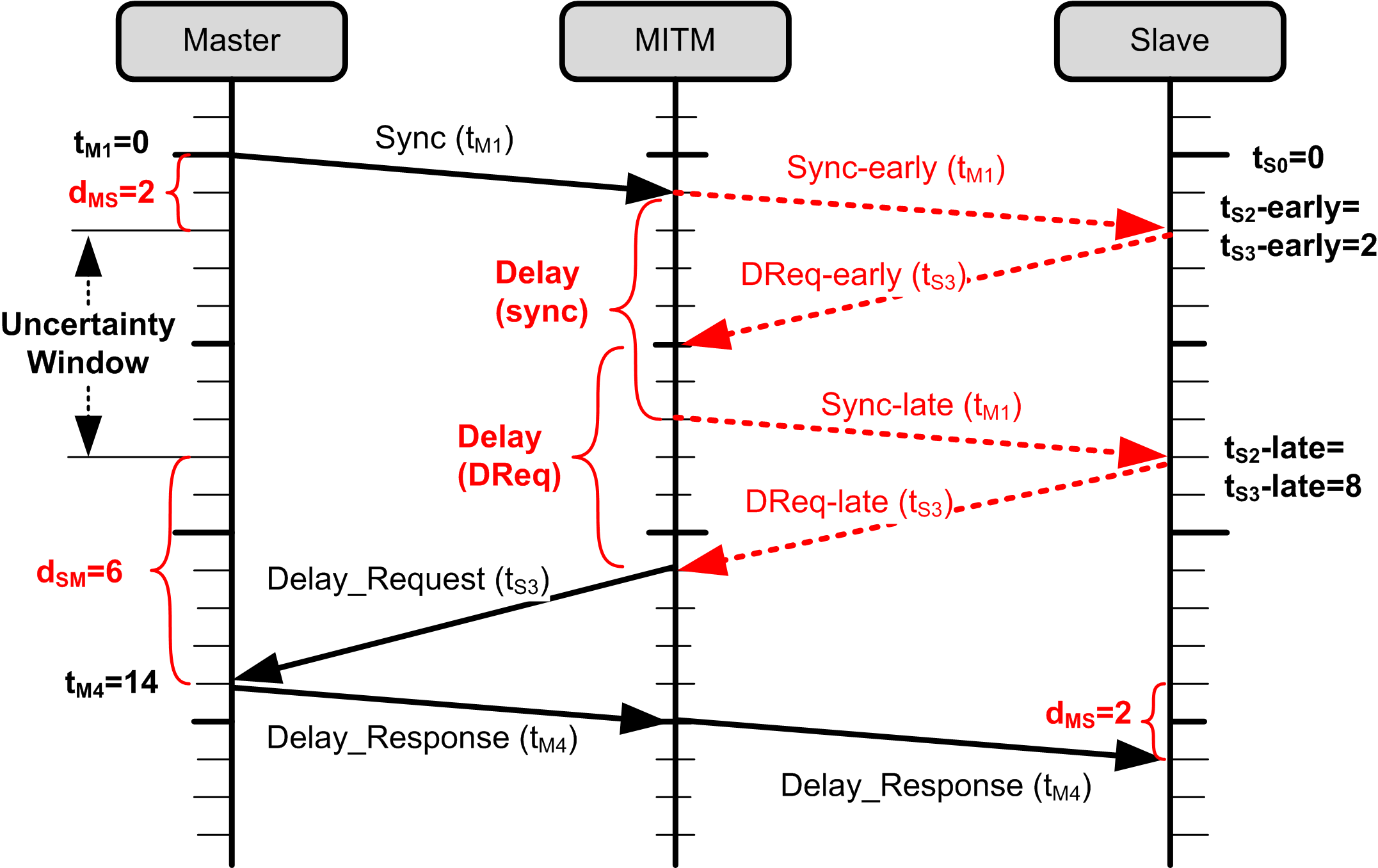}
	\caption{Offset uncertainty bound calculation using one-way delay limits and unsynchronized clocks.}
	\label{fig:link_constraints-rtd}
\end{figure}

For computing guaranteed offset bounds using $d_{min}^{MS}$ and $d_{min}^{SM}$, the slave first measures the \gls{rtd} according to Eq.~\eqref{eq:rtd}. 
In Fig.~\ref{fig:link_constraints-rtd} the forward communication path from master to slave has a known minimum \gls{owd} of $d_{min}^{MS}=2$ and the reverse path a minimum \gls{owd} of $d_{min}^{MS}=6$ time units. 
Relying on the minimum delay constraints, whenever the slave receives the master's \sync message it knows that the master assigned timestamp $t_{M1}$ at least $d_{min}^{MS}=2$ time units earlier then the slave's reception timestamp $t_{S2}$. The slaves also knows when sending its \delayreq message that the master will receive it and assign timestamp $t_{M4}$ at least $d_{min}^{MS}=6$ time units later then slave time $t_{S3}$. 
Using this knowledge and the timestamp $t_{M4}$ it received in the master's \delayresp message, the slave can rely on the inequalities~\eqref{eq:offsetuneqa} and \eqref{eq:offsetuneqb} to hold true.

Assuming an (unknown) clock offset $\mathit{offset}$ between the slave and master clocks, the slave can define the causal ordering of timestamps using the inequalities Eq.~\eqref{eq:offsetuneqa} and Eq.~\eqref{eq:offsetuneqb}.   
All terms except the offset being known, by reordering the inequalities the slave can bound its clock offset after a clock synchronization interval by Eq.~\eqref{eq:offsetuneqa1}. 

\begin{subequations}
	\begin{eqnarray}
	t_{M1} +  d_{min}^{MS} + \mathit{offset} \leq t_{S2}\label{eq:offsetuneqa}\\
	t_{M4} \geq t_{S3} + d_{min}^{SM} - \mathit{offset}\label{eq:offsetuneqb}\\
	t_{S3} - t_{M4} + d_{min}^{SM} \leq \mathit{offset} \leq t_{S2} - t_{M1} - d_{min}^{MS}\label{eq:offsetuneqa1}
	\end{eqnarray}
\end{subequations}

Applying Eq.~\eqref{eq:offsetuneqa1} to the scenario in Fig.~\ref{fig:link_constraints-rtd} yields bounds of $\interval{-6}{0}$ for the early case $t_{S2}=t_{S3}=2$ and $\interval{0}{6}$ for the late case $t_{S2}=t_{S3}=8$, which maps to the uncertainty window depicted in the figure. 
In order to obtain a clock offset bound that centers around \num{0} (despite asymmetric communication path and delay attacks), we suggest replacing clock offset calculation from Eq.~\eqref{eq:offset} by Eq.~\eqref{eq:offsetowd}, which basically averages the minimum and maximum offset, assuming that the uncertainty (i.e., difference between the measured \gls{rtd} and the sum of the minimum \glspl{owd} $d_{min}^{MS}$ and $d_{min}^{SM}$) affects the forward or reverse link with equal probability. In Fig.~\ref{fig:link_constraints-rtd} the average offset is mapped to the center of the marked uncertainty window. This yields a symmetrical interval for guaranteed clock offset bounds that formally satisfy  Eq.~\eqref{eq:offsetboundsimple} with $\gls{rtd} = t_{M4} - t_{M1}$.  
For the scenario in Fig.~\ref{fig:link_constraints-rtd}, the new offset calculation results in the same size of the uncertainty window (i.e., \num{6}) but the offset within $\interval{-3}{3}$ centered around \num{0} (according to Eq.~\eqref{eq:offsetboundsimple}). 
The knowledge of the physical delay therefore allows a stricter bound of the guaranteed clock offset compared to the case without knowledge on the \glspl{owd} presented in Fig.~\ref{fig:link_constraints-theoretical}. 
In this way, the adversary's degree of freedom in manipulating clock synchronization is effectively decreased and the guaranteed bounds on the clock offset are improved. 

\begin{equation}
    \begin{split}
	\mathit{offset} & = \frac{\mathit{offset_{low}} + offset_{high}}{2}\\
	& = \frac{t_{S2} - t_{M1} - d_{min}^{MS} + t_{S3} - t_{M4} + d_{min}^{SM}}{2}\label{eq:offsetowd}
    \end{split}	
\end{equation}

\begin{equation}
	- \frac{\mathit{RTD} - d_{min}^{MS} - d_{min}^{SM}}{2} \leq \mathit{offset} \leq \frac{\mathit{RTD} - d_{min}^{MS} - d_{min}^{SM}}{2}\label{eq:offsetboundsimple}
\end{equation}

While the clock offset can be guaranteed after a clock synchronization interval (Eq.~\eqref{eq:offsetboundsimple}), the question about the clock offset that can be guaranteed for a particular system remains open as an attacker could still increase the \gls{rtd} arbitrarily (and therefore manipulate the clock offset). 
To prevent such clock offset manipulation, the \gls{rtd} that is accepted needs to be restricted. 
Intuitively, the lower the maximum \gls{rtd} that is accepted ($\mathit{RTD}_{max}$), the tighter the bound on the clock offset that can be guaranteed, but also the higher the probability of clock synchronization intervals to be discarded during unfavorable network conditions. 

In order to derive clock offset bounds for a system, a time $T_I$ needs to be defined that represents the maximum time interval between any two consecutive clock synchronization intervals with $\mathit{RTD} \leq \mathit{RTD}_{max}$. 
Between two consecutive clock synchronization intervals, the slave clock drifts at most $T_I \cdot \abs{\rho}$, with $\rho$ being the maximum relative clock drift of slave and master clocks. 
Generally one can say that the smaller the network the smaller $\mathit{RTD}_{max}$, the better the network operations the smaller $T_I$, and the better the clocks the smaller $\rho$. 
Eq.~\eqref{eq:offsetboundsystem} shows the guaranteed clock offset bounds for a system. 

\begin{equation}
\begin{split}
- \frac{\mathit{RTD}_{max} - d_{min}^{MS} - d_{min}^{SM}}{2} - T_I \cdot \abs{\rho} \leq \mathit{offset}\\
 \mathit{offset} \leq \frac{\mathit{RTD}_{max} - d_{min}^{MS} - d_{min}^{SM}}{2} + T_I \cdot \abs{\rho}\label{eq:offsetboundsystem}
\end{split}
\end{equation}

While the relative clock offset can be bounded, it needs to be stressed that high-precision clock synchronization requires significantly tighter clock synchronization guarantees than can be provided by the bounds in Eq.~\eqref{eq:offsetboundsimple} and~\eqref{eq:offsetboundsystem}. 
Deterministic networks might help as they provide guarantees on the maximum \gls{rtd} and delay variation, but they depend on a precise notion of time themselves in the first place. 

It is worth noting that slaves can apply the algorithm that has been described above for the \sync and \delayreq message round-trip for the \delayreq \delayresp message exchange, too. The \gls{ptp} protocol does not evaluate the \delayresp message's delay as part of its operation, which is why the \sync and \delayreq message exchange is the preferred one for clock bounding. However, the slave could acquire the \delayresp receiving timestamp and use it to compute clock bounds for \delayreq \delayresp. One possible application is, for instance, the detection of selective \sync message delays conducted by a resource-limited adversary. The clock offset bounds of the \sync \delayreq round-trip being substantially larger (inaccurate) than the one of \delayreq \delayresp could be a strong indication of an ongoing selective delay attack. Still, the general observation holds true: none of these methods can protect against an asymmetrical link delay attack that an adversary with unlimited resources can conduct.

	\section{Clock Synchronization: Either Precise or Secure}
	\label{sec:conjecture}

The outcome of Section~\ref{sec:asymmattack} that an attacker can manipulate offset correction by $\left [\frac{-\delta_1}{2}, \frac{\delta_2}{2} \right]$ through link asymmetry raises the question whether a clock synchronization protocol can be designed that handles link asymmetry such that its offset calculation remains unaffected in an adversarial setting. 
We argue that constructing such protocol is impossible as delay cannot be distinguished from clock offset. 
Main reason is that link asymmetry can only be measured with synchronized clocks, and it can not be guaranteed that the link asymmetry is constant (especially in an adversarial setting). 
Synchronized clocks would be required in the first place to measure link asymmetry in order to have secure clock synchronization after all, which is circular reasoning. 

If we suppose an oracle to exist that can instantaneously read the clocks of master and slave, the oracle has knowledge of the real clock offset (according to Eq.~\eqref{eq:offset}) and is not influenced by link asymmetry. 
The clock offset measured by the slave also includes the link asymmetry (Eq.~\eqref{eq:measuredoffset}), however. 

\begin{equation}
\mathit{offset}_{measured} = \mathit{offset}_{real} + \frac{\delta_2 - \delta_1}{2}
\label{eq:measuredoffset}
\end{equation}

If the oracle analyzes two consecutive offset measurements (at time $i$ and $i+1$) during which clock offset was not corrected, it would observe that the difference between the offsets measured at time $i$ and at time $i+1$ consists of two distinct parts: (1) the change in the real offset ($\mathit{offset}_{real_{i+1}} - \mathit{offset}_{real_i}$) that is a result of the relative clock drift between slave and master, and (2) the change in the link asymmetry that is determined by delay variation ($\frac{\delta_{2_{i+1}} - \delta_{1_{i+1}}}{2} - \frac{\delta_{2_i} - \delta_{1_i}}{2}$) such as network jitter for example. 

The relative clock drift, which determines the change of the real offset, depends on the quality of the physical oscillators used and typically ranges from $10^1$ to $10^{-6}$ ppm. 
The delay variation is highly indeterministic and depends on various factors such as the current network load as well as the quality of the network and its components. 
Delay variation typically ranges from $10^5$ to $10^2$ ppm. 
Important to note is that delay variation is by several orders of magnitudes larger than the relative clock drift, and that only an oracle could distinguish the change of the real clock offset from the change of the link asymmetry. 
For master and slave, however, they are indistinguishable as only the sum of the change can be observed in terms of measured clock offset. 
This means that an attacker can exploit this indistinguishability to conduct and hide an asymmetric link delay attack. 
As soon as a clock synchronization protocol aims to achieve high precision, it needs to entail delay measurements in order to compensate for delays. 
And this delay compensation mechanism is susceptible to asymmetric links because link delay variation cannot be separated from clock drift. 

The other issue is the direction of messages. 
To conduct an asymmetric link delay attack, the attacker only needs to know the direction of messages, which is tied to the master and slave roles because it is always the clock offset of the slave relative to the master that is calculated. 
Delaying messages in one direction has the inverse effect on clock synchronization than delaying messages in the reverse direction (as the slave clock should be synchronized to the master and not the other way around). 
For this reason we conclude that no clock synchronization protocol can be designed that is precise and prevents delay attacks entirely (even when messages are obfuscated in terms of length and timing and cover traffic exists). 

If clock synchronization protocols can be either high-precision or secure against delay attacks, then applications must not rely on a precise notion of time when employing untrusted communication networks --- if applications rely on a precise notion of time, then they must only be executed in trusted environments. 
This conclusion is (not limited to but) especially important to critical infrastructures. 
 	
	\section{Conclusion}
	\label{sec:conclusion}

In this paper, we focused on attacks against clock synchronization protocols in the time domain, which means that protocol message content is not altered and only the timing of messages is changed. 
One assumption is that the attacker is in a privileged network position\footnotemark. 
We first conducted a statistical traffic analysis of \gls{ptp} and identified properties of \gls{ptp} traffic with regard to timing, packet length, and packet directions. 
We showed that these properties can be used to identify \gls{ptp} messages in encrypted traffic in order to conduct selective message delay attacks. 

\footnotetext{While we assume the difficulty of gaining such privileged network position is within the power of an attacker who attacks critical infrastructures, we think that asymmetric link delay attacks specifically might be conducted even from non-privileged network positions by influencing the queues of network devices in a particular direction (for example by sending an excessive number of packets). }

We explored various countermeasures to mitigate selective message delay attacks. 
The first set of countermeasures aims to obstruct traffic analysis. 
To this end, \gls{ptp} can be modified in a way that randomizes the timings and the use of packet length padding, although such modification may have a negative impact on the clock synchronization's precision. 
Security, nevertheless, depends on the existence of suitable cover traffic, which leads to the field of traffic obfuscation. 
Furthermore, strict replay protection should be activated if possible to minimize the impact of the attack (and to make the attack easier to detect as the packet loss rate increases). 

Then we introduced asymmetric link delay attacks. 
While asymmetric link delay attacks have potentially lower impact on clock synchronization's precision, we found that they are fundamentally tied to the goal of high-precision. 
Bounding the uncertainties of the clock offset by applying knowledge of the physical parameters of the communication path (i.e., limiting \glspl{owd}) ensures that individual messages cannot be delayed arbitrarily. 
Until now, network-based clock synchronization protocols asked for deterministic delays, and delays could be either symmetric or have a known asymmetry to be compensated by the \gls{ptp} configuration. 
The results show that knowledge of the underlying communication networks is essential to mitigate delay attacks and to safeguard maximum guaranteed bounds on clock offset. 
Nevertheless, asymmetric link delay attacks can only be mitigated but not be prevented entirely. 

We argue that no high-precision clock synchronization protocol can exist that prevents asymmetrical link delay attacks entirely because of the delay compensation mechanism that is required to achieve high-precision. 
In adversarial settings, an attacker can manipulate the delay variation in such a way that links become asymmetric and clock offset calculation is impaired maliciously since clock drift and delay variation cannot be distinguished. 
This implies that clocks synchronization cannot be arbitrarily precise while maintaining security against delay attacks. 
This finding contradicts the common belief that clocks synchronization over untrusted networks can be secured by encryption and authentication methods, while improving precision. 
Given the results from this paper, we argue the contrary: clock synchronization can either be precise or secure against delay attacks (but not both!). 

Delay attacks are an inherent threat for high-precision clock synchronization since the times when messages are sent and received have an actual effect on the precision of clock synchronization and even small differences can have a large impact. 
The impact of those delay attacks can only be bounded but those attacks limit the precision of clock synchronization, nevertheless. 
Practically achievable limits are certainly stricter than some critical infrastructure applications assume today. 
Those infrastructures are supposed to improve specific areas but also introduce a new attack vector by their strict dependency on a precise notion of time.

	\printbibliography[heading=bibintoc]

\end{document}